\documentclass[conference]{IEEEtran}
\IEEEoverridecommandlockouts
\usepackage{xcolor}
\usepackage{cite}
\usepackage{amsmath,amssymb,amsfonts}
\usepackage{algorithmic}
\usepackage[hyphens]{url}
\usepackage{float}
\usepackage{comment}
\usepackage{tabularx}
\usepackage{multirow}
\usepackage{pdflscape}
\usepackage{booktabs}
\usepackage{longtable}
\usepackage{tabularray}
\usepackage[hidelinks]{hyperref}
\usepackage{graphicx}
\usepackage{adjustbox}
\usepackage[OT1]{fontenc} 
\usepackage{balance}
\usepackage{rotating} 

\usepackage{array}
\newcolumntype{L}[1]{>{\raggedright\let\newline\\\arraybackslash\hspace{0pt}}m{#1}}
\usepackage{textcomp}
\usepackage{xcolor}
\def\BibTeX{{\rm B\kern-.05em{\sc i\kern-.025em b}\kern-.08em
    T\kern-.1667em\lower.7ex\hbox{E}\kern-.125emX}}
    
\begin{document}

\title{Global Public Sentiment on Decentralized Finance: A Spatiotemporal Analysis of Geo-tagged Tweets from 150 Countries 
}

\author{
  \IEEEauthorblockN{%
    Yuqi Chen\IEEEauthorrefmark{3}
    Yifan Li\IEEEauthorrefmark{4}
    Kyrie Zhixuan Zhou\IEEEauthorrefmark{5}
    Xiaokang Fu\IEEEauthorrefmark{2}
    Lingbo Liu\IEEEauthorrefmark{2}\\
    Shuming Bao\IEEEauthorrefmark{6}
    Daniel Sui\IEEEauthorrefmark{7}
    Luyao Zhang\IEEEauthorrefmark{1}}

    \IEEEauthorblockA{%
    \IEEEauthorrefmark{3}Peking University, Beijing, China\\
    \IEEEauthorrefmark{4}University of Toronto, Toronto, Canada\\
    \IEEEauthorrefmark{5}University of Illinois at Urbana-Champaign, Champaign, USA\\
    \IEEEauthorrefmark{2}Center for Geographic Analysis, Harvard University, Cambridge, USA\\
    \IEEEauthorrefmark{6}China Data Institute, Ann Arbor, USA\\
    \IEEEauthorrefmark{7}Virginia Tech, Blacksburg, USA\\
    \IEEEauthorrefmark{1}Data Science Research Center and Social Science Division, \\ Duke Kunshan University, Suzhou, China
  }
\thanks{\textsuperscript{\textasteriskcentered}The corresponding author: Luyao Zhang (lz183@duke.edu). Yuqi Chen and Yifan Li contributed equally to this research. \textbf{Acknowledgments.} Luyao Zhang is supported by the National Science Foundation China (NSFC) on the project entitled “Trust Mechanism Design on Blockchain: An Interdisciplinary Approach of Game Theory, Reinforcement Learning, and Human-AI Interactions (Grant No. 12201266). This work is partially sponsored by the National Science Foundation in the United States (Grant No. 1841403). We sincerely thank the participants and organizers of the Invited Talk at the Department of Land Surveying \& Geo-Informatics (LSGI), The Hong Kong Polytechnic University, Hong Kong, China (August 12, 2024); the Invited Seminar at Financial Technology Thrust, Society Hub, The Hong Kong University of Science and Technology (Guangzhou), Guangzhou, China (November 8, 2024); and the KNIME Business Hub for Spatiotemporal Data Science: Research, Teaching, and Business Applications at the Center for Geographic Analysis (CGA), Harvard University (December 16, 2024) for valuable discussions that enriched this work. Special thanks to Prof. Campbell Harvey, Prof. Xintao Liu, Prof. Xuechao Wang, and Prof. Leifu Zhang for their insightful comments and guidance. We are grateful to Devika Jain and Weihe Wendy Guan for their generous guidance on utilizing the datasets from the Harvard Center for Geographic Analysis Geotweet Archive: \url{https://doi.org/10.7910/DVN/3NCMB6}.} 
}

\maketitle


\maketitle

\begin{abstract}
Blockchain technology and decentralized finance (DeFi) are reshaping global financial systems, yet existing research often overlooks the spatial distribution of public sentiment and its economic and geopolitical determinants. To address this gap, we analyzed over 150 million geo-tagged, DeFi-related tweets from 2012 to 2022, drawn from a comprehensive dataset of 7.4 billion tweets. Employing sentiment scores generated by a BERT-based multilingual classification model, we integrated these tweets with economic and geopolitical data to construct a multimodal dataset. We utilized techniques such as sentiment analysis, spatial econometrics, clustering, and topic modeling to elucidate the drivers of engagement and sentiment. Our results reveal significant global variations in attention to DeFi and the sentiments expressed, with economic development emerging as a pivotal influence, particularly after 2015. Geographically weighted regression analysis confirms GDP per capita as a key predictor of the proportion of DeFi-related tweets, with its influence intensifying following notable increases in cryptocurrency values, such as Bitcoin. Wealthier nations generally exhibit more active engagement in DeFi discourse, while the lowest-income countries, displaying high attention and positive sentiments, frequently frame DeFi in terms of financial security and potential for sudden wealth. In contrast, middle-income countries contextualize DeFi within social and religious narratives, whereas high-income countries predominantly regard it as a speculative instrument or a form of entertainment. This study not only advances interdisciplinary research in computational social science and finance but also contributes to open science by releasing our dataset and analytical code on GitHub, and by providing a non-code workflow on the KNIME platform. These resources empower a diverse range of scholars to investigate DeFi adoption and sentiment, thereby informing policymakers, regulators, and developers in promoting financial inclusion and fostering responsible DeFi engagement worldwide.
\end{abstract}


\section{Introduction}

The emergence of blockchain technology, cryptocurrencies, and non-fungible tokens (NFTs) has fundamentally reshaped our understanding of finance, trust, and decentralized systems~\cite{monrat2019survey,zhang2023design}. Among these innovations, decentralized finance (DeFi) has become a disruptive force, challenging traditional financial intermediaries by providing open, permissionless, and algorithm-driven financial services~\cite{harvey2021defi,zetzsche2020decentralized}. As DeFi continues to expand and integrate into the global economy, it is essential to understand the factors driving its adoption and impact.

One critical determinant of DeFi adoption is public sentiment, which significantly influences how emerging financial technologies are embraced and incorporated into the broader economy~\cite{zhuo2024texts,okat2024trust}. Research on the adoption of other emerging technologies, such as artificial intelligence (AI), has demonstrated that positive sentiment fosters higher adoption rates and investment flows, whereas negative perceptions contribute to hesitancy, regulatory scrutiny, and reduced participation~\cite{lui2022impact,polisetty2024determines}. Social media platforms, particularly Twitter, serve as valuable data sources for tracking public opinion and sentiment, providing real-time insights into market trends and user perceptions~\cite{kuziemski2020ai}. Previous studies on sentiment toward decentralized finance have primarily examined its role in forecasting the value of cryptocurrencies~\cite{Canayaz2023} and NFTs~\cite{Zhang2023}, as well as the differing attitudes toward DeFi among various groups, such as Key Opinion Leaders (KOLs) and general users~\cite{Xiao2023}. However, there remains a notable gap in research from a spatial analysis perspective. Few studies have investigated how public attention and sentiment toward DeFi vary across different countries and regions. Exploring these variations is crucial for understanding global disparities in the perception and engagement with decentralized financial technologies, which would, in turn, affect the adoption of the emerging technology DeFi and thus influence economic development. The lack of spatially oriented research represents a missed opportunity to assess how economic, regional, and geopolitical factors shape public discourse and sentiment regarding emerging digital financial systems.

Our research is unique in its spatial analysis of geographically tagged Twitter data, which enables the integration of public attention and sentiment across different timeframes and countries. This approach allows us to identify how regional economic conditions and geopolitical contexts influence public sentiment toward DeFi, offering new insights into the global adoption of decentralized financial technologies.

As an initial exploratory step to pave the way for further causal hypothesis testing, this study aims to investigate the following research questions (RQs):

\textbf{RQ1:} Are there statistically significant variations in public attention and sentiments expressed on social media toward decentralized finance across different countries and regions? If so, what are the key differences?

\textbf{RQ2:} Are spatial variations in public attention and sentiment significantly associated with economic factors, and how do these dynamics evolve over time?

\textbf{RQ3:} What specific topics are countries and regions most likely to discuss when engaging with decentralized finance, and how do these topics differ across clusters formed based on economic, public attention, and sentiment differences?

To address RQ1, we compiled a dictionary of 152 multilingual keywords to identify tweets related to decentralized finance, including blockchain, NFTs, and cryptocurrencies. By aggregating data, we computed sentiment scores and the proportion of decentralized finance-related tweets across approximately 150 countries from 2012 to 2022. Our preliminary findings indicate significant global variations in both the level of attention to decentralized finance—particularly when controlling for general engagement on Twitter—and the sentiments expressed toward these emerging topics.

To address RQ2, we introduced GDP per capita as a proxy for the economic development of different countries. Through exploratory analysis, we examined the correlations among GDP per capita, tweet proportion, and sentiment score to investigate how economic factors influence public attention and sentiment towards decentralized finance across various countries. We then conducted a geographically weighted regression analysis between GDP per capita and the proportion of tweets concerning decentralized finance for each year. Our findings reveal that GDP per capita becomes a significant predictor of tweet proportion from 2015 onwards, highlighting the role of economic development in driving social media engagement with new technologies. Controlling for general engagement on Twitter, we find that more economically developed countries show higher levels of engagement on DeFi, while less developed nations exhibit lower levels. To assess changes over time, we performed a Pearson correlation analysis between GDP per capita and tweet proportion for each year, followed by Fisher's z transformation of the correlation coefficients. The results indicate that the influence of GDP on attention to decentralized finance fluctuates over time, often aligning with changes in the price of digital currencies, such as Bitcoin. Notably, after significant increases in Bitcoin's value, the impact of GDP on attention levels tends to strengthen, underscoring the sensitivity of economically developed countries to shifts in economic value. Furthermore, countries with lower economic development but high engagement in decentralized finance often display the highest sentiment levels.

To address RQ3, we applied clustering analysis, data visualization, and topic modeling to examine tweets from each of the three identified groups of countries. The clustering analysis categorized countries based on GDP per capita, public attention toward decentralized finance, and sentiment expressed on social media, yielding an optimal Silhouette Score of 0.312. While higher-income countries generally exhibited greater public attention to decentralized finance, lower-income countries presented an anomaly, demonstrating the highest levels of engagement and the most positive sentiment, whereas middle-income countries displayed the lowest levels of both attention and sentiment. Furthermore, with the exception of a few outliers such as China and South Korea, countries where cryptocurrency is legally recognized tend to have higher GDP per capita, and the legal status of cryptocurrencies follows a pattern in which economically advanced nations are more likely to permit cryptocurrency trading. The subsequent topic analysis provides insight into this anomaly, revealing that discussions in lower-income countries are more emotionally charged and frequently frame decentralized finance as a means of achieving financial security or sudden wealth, as indicated by terms such as “luck,” “hope,” and “promote.” In contrast, middle-income countries emphasize the intersection of decentralized finance with social and religious contexts, while high-income countries primarily regard cryptocurrencies as financial instruments for investment or entertainment.
\par
The rest of the paper is organized as follows. Section~\ref{sec:related work} discusses related work. Section~\ref{sec:data} elaborates on the data sources. Section~\ref{sec:methods} outlines the methodology of analysis. Section~\ref{sec:results} presents the results. Section~\ref{sec:discussion} discusses the limitations and future research. Finally, Section~\ref{sec:conclusion} concludes.

\textbf{Data and Code Availability Statement}: We have made our dataset and code publicly available as open-source on GitHub\footnote{\url{https://github.com/yukiyuqichen/GeoBlockchain}}



\section{Related Work}
\label{sec:related work}
Our research contributes to the realm of decentralized finance and social media studies.

\subsection{Decentralized Finance}
The up-to-date prosperity of decentralized finance derived from the subversive reform in digital finance, a ramification of finance that tenders a series of financial services on the Internet through digital devices or platforms \cite{ozili2022decentralized}. The operation of decentralized finance is underpinned by the public database called blockchain, which records unalterable data publicly and digitally\cite{dinh2018untangling, tseng2020blockchain}. It is also described as a digital ledger that provides duplication and storage of transactions across many computer networks \cite{adams2018blockchain, riva2020happens}. By the means of employing blockchain, decentralized finance erodes and challenges the power of conventional intermediaries. Different from centralized finance, it eliminates the negative impacts of the failure of intermediaries, reduces the cost of transactions, and reduces the credit risk, which usually occurs when default happens \cite{atlam2018blockchain, swan2017anticipating, garg2021measuring}. 

Although decentralized finance generates benefits economically, its risks cannot be neglected and are imperative to apprehend. First, risk exists in the transaction procedures due to the coding error in the process of forming the smart contract. The coding error generates the vulnerabilities that engender the situation of stealing funds through smart contracts by attackers \cite{schar2021decentralized}. Second, the unpredictable legal liability may jeopardize the interests of users due to their restricted comprehension of the smart contract. They struggle with understanding the codes and assessing the security properly. Therefore, they will mistakenly sign the comprised contract that they are exposed to certain legal risks \cite{schar2021decentralized}. Third, there is a risk of data stealing. Usually, users use admin keys to control the smart contract and conduct emergency shutdowns. If the key holder fails to place their key securely, the others can steal the keys to change or compromise the smart contract \cite{schar2021decentralized}. Another risk is the reliance on external data. The DeFi applications rely on external data when the data of the smart contract is not available in the blockchain. The data quality and data availability can be questioned if they severely depend on the external data \cite{schar2021decentralized, 9154123}. Last but not least, the decentralized characteristics facilitate the development of illicit activities through decentralized finance applications. It furnishes lawbreakers a safe and effective venue to commit crimes, such as laundering money and funding criminal activities \cite{schar2021decentralized}.

Putting the development of decentralized finance into a global perspective, the number of applications, protocols, users, and research related to decentralized finance skyrocketed in recent years, indicating people's growing attention and interest in this field and highlighting the importance of the notion of decentralized finance in today's global environment \cite{ozili2022decentralized, meyer2022decentralized}. Blockchain can be considered as the most far-reaching technology in the next decades, especially for the financial inclusion of the unbanked people \cite{yaga2019blockchain, abdulhakeem2021powered, demirgucc2020global}. According to Meyer's \cite{meyer2022decentralized}contribution to the DeFi literature review, up to 2022, 55 out of 83 DeFi articles were published in 2020 and the first half of 2021. From 2018 to 2021, there is a 6900 per cent increment in the users of decentralized finance protocols, rising from 3,000 to over 210,000 users. What's more, the total value locked in DeFi exceeds 83 billion USD, representing an increase of over 60,000 percent. The increased global academic interests, protocol users and total value further elaborates the importance of DeFi worldwide and demonstrates its revolutionary and exceptional power. 

Although multifarious articles and research exist and are published nowadays, we discover a gap in the research of the global perspective of decentralized finance that fails to emphasize the differentiation of public attention and sentiments among countries. With the geo-tagged data, our article contributes to fulfilling the gap by conducting spatial analysis to illustrate and unfold the global imbalances in public attention and sentiments toward decentralized finance. Another contribution is to find out which factors impact the public attention and sentiment towards decentralized finance among countries through cluster analysis and Geographically Weighted Regression. 

\subsection{Social Media}
Social Media plays an important role in transforming people's life, providing a digital platform for people to exchange ideas, communicate with each other, access the updated information and news, seek advice, engage in social activities, etc \cite{carr2015social,amedie2015impact, siddiqui2016social,edosomwan2011history}. It has profound and prominent impacts on different areas, from business to criminal activities \cite{amedie2015impact, edosomwan2011history}. For example, the sentiment information shown in social media contains statistically significant indications of the future prices of the S \&\ P500 index and a limited set of stocks, demonstrating the prediction power of social media on stock markets \cite{zheludev2014can}. 

Our research primarily examines the influence of social media on decentralized finance. Previous studies have largely focused on how information shared on social media, particularly tweets, impacts the prices, trading volumes, and volatility of cryptocurrencies, with findings indicating statistically significant relationships \cite{rouhani2020crypto, hamza2020effect, awad2022impact, su2022social}. Differing from these studies, our contribution lies in utilizing geotagged information and sentiment scores from tweet data to illustrate the imbalanced patterns of public attention and sentiment towards decentralized finance across different countries. Additionally, we employ topic modeling and cluster analysis to identify keywords that highlight notable topics and daily discussions related to decentralized finance among countries within distinct clusters.

\section{Data}
\label{sec:data}


\subsection{Geotagged Tweets}
In this study, we utilized the Harvard University Twitter Archive v2.0 \cite{lewisHarvardCGAGeotweet2023}, which comprises geotagged, multilingual tweets from around the world, spanning the period from January 2010 to June 2023. The dataset includes various fields, such as timestamps, text, location, language, and more. A representative example of the dataset is provided in Table \ref{tab:sample} in Appendix \ref{sec: Data}.
From a total of 7.4 billion tweets in this archive, we extracted 15,020,385 tweets containing specific keywords related to blockchain, NFTs, and cryptocurrencies in different languages. The keywords are provided in Table~\ref{tab:dictionary}.

The geotagged tweet data employs two attributes, Coordinates and Place, to represent the spatial location of a tweet. Each attribute is discussed as follows \cite{chai2023twitter}:

\textbf{Coordinates}: This attribute specifies the geographic location of the tweet as provided by the user or client application. The coordinates array is formatted in GeoJSON with longitude preceding latitude.

\textbf{Place}: This attribute signifies that the tweet is linked to a specific place, generally a town name as defined by the user. Twitter establishes a bounding box based on this information, using its centroid as the coordinates if actual GPS data is unavailable. The spatial error is estimated by calculating the radius of a circle derived from this bounding box.

Key fields for location signatures are listed as follows:

\textbf{Latitude and Longitude}: Every tweet includes these fields, which are either derived from the Twitter Coordinates object or calculated using the centroid of the bounding box from the Place object.

\textbf{GPS}: This flag indicates whether the tweet's coordinates are sourced from GPS data or the Place name-based bounding box. If the coordinates are from GPS, this field is marked “Yes”; otherwise, it is marked “No”. When both sources are available, the GPS coordinates take precedence. This distinction helps determine if the coordinates are actual GPS data or derived from a place name.

\textbf{Spatial Error}: This field, which is crucial for interpreting tweet locations, estimates the horizontal error in meters. A 10m error is assumed for GPS coordinates, while the error for place name-based coordinates is calculated as the radius of a circle with the area of the bounding box.

\subsection{Twitter Sentiment Geographical Index}
We obtained the sentiment scores of geotagged tweets from the Twitter Sentiment Geographical Index (TSGI) Dataset \cite{chaiTwitterSentimentGeographical2023} (\url{https://gis.harvard.edu/twitter-sentiment-geographical-index-tsgi-dataset-global-high-frequency-dataset-monitoring}). By applying a BERT-based multi-lingual sentiment classification model to the comprehensive archive of 7.4 billion geotagged tweets, they constructed a high-frequency multi-year database that has global coverage and enables the evaluation of subjective well-being (SWB) in 163 countries and regions for one decade since 2012.

\subsection{World Development Indicators}
We utilized data from the World Bank's World Development Indicators (WDI), specifically focusing on the GDP per capita metric. Considering that our Twitter data spans from 2012, we calculated the average GDP per capita for different countries from 2012 to 2022, providing an accurate representation of each country's economic status during this decade and allowing us to analyze how economic factors might influence public attention and sentiments on decentralized finance in different regions.

\subsection{Cryptocurrency Regulation}
We incorporated the data from `Cryptocurrency Regulation Tracker' (\url{https://www.atlanticcouncil.org/programs/geoeconomics-center/cryptoregulationtracker/}), which focuses on the regulatory landscape of cryptocurrencies across 60 countries.
Each country is assigned one of the following regulatory statuses: legal (where all activities are permitted), partial ban (where one or more activities are not permitted), and general ban (where all activities are limited).






\section{Methodology}
\label{sec:methods}

\subsection{Spatial Autocorrelation Analysis}
Spatial Autocorrelation depicts the relationship between the contiguous spatial units, where each one of them is coded with the realization of a single variable \cite{getis2009spatial}. 

Realizing that our variables include spatial information, it is mandatory to check spatial autocorrelation before we conduct the regression analysis. If our variables suffer from spatial autocorrelation, we have to adjust our regression method from simple linear regression to geographically weighted regression. 

To test the Spatial Autocorrelation for our continuous variables (GDP per capita, tweet proportion, and sentiment), we apply Moran's I method. The general formula of Moran's I statistic is shown below \cite{tiefelsdorf1995exact}.
\begin{equation}
\begin{split}
I = {n \over A}{{y^{T}MCMy}\over{y^{T}MMy}}, \\
n = 1^{T}1 \\
A = 1^{T}C1 \\
M = I-X(X^{T}X)^{-1}X^{T}
\end{split}
\end{equation}
Where \(y\) is the vector of dependent variables, \(C\) is the nonstochastic connectivity matrix defined by the underlying spatial structure, \(X\) is the nonstochastic regression matrix of independent variables (\(n\)$\times$\(p\)), the constant column \(1\) is the intercept of the regression equation, \(p\) is the number of independent variables, \(n\) is the number of spatial entities, \(A\) is the overall connectivity in the system, and \(M\) is the projection matrix. 

For our categorical variables clusters, we introduce the joint count statistic \cite{moran1948interpretation}, which is the spatial analysis method for categorical variables in autocorrelation. In joint count statistics, we have binary variables \(X_{i} \in {0,1}\) in the distribution of N spatial sites, where the neighborhood relations between regions \(i\) and \(j\) are represented in matrix \(W_{ij}\).
\begin{equation}
    W_{ij}=
\begin{cases}
    1, & \text{if}\ i \text{ is the neighbor of } j \\
    0, & \text{otherwise}
\end{cases}
\end{equation}
The statistics is defined as \cite{cliff1981spatial, dale2014spatial}:
\begin{equation}
\begin{split}
    J = J_{BB} + J_{BW} + J_{WW}, \\
    J_{BB} = {1 \over 2}{\sum_{ij, i \neq j}} W_{ij}X_{i}X_{j} \\
    J_{BW} = {1 \over 2}{\sum_{ij, i \neq j}} W_{ij}(X_{i}-X{j})^{2} \\
    J_{WW} = {1 \over 2}{\sum_{ij, i \neq j}} W_{ij}(1-X_{i})(1-X_{j}) \\
    J = {1 \over 2}{\sum_{ij, i \neq j}} W_{ij}
\end{split}
\end{equation}

\subsection{Geographically Weighted Regression}
\label{Geographically Weighted Regression}
Due to the spatial autocorrelation in our variables, we introduce the Spatial Error Model (SEM) method, a linear regression model with the spatial autoregressive error term \cite{zhang2015improved}. The model is shown as the following \cite{saputro2019spatial}.
\begin{equation}
\begin{split}
Y = X\beta + \mu \\
\mu = \lambda W \mu + \epsilon \\
\epsilon \sim N(0,\sigma^{2}I)
\end{split}
\end{equation}

Where \(Y\) is the vector of dependent variables (\(n \times 1\)), \(X\) is the matrix of the independent variables (\(n \times k\)), \(\beta\) is the vector of regression coefficients (\(k \times 1\)), \(\mu\) is the vector of error term(\(n \times 1\)), \(\lambda\) is the autoregressive coefficient ranged from -1 to 1, \(I\) is the identity matrix, and \(W\) is the matrix of standardized spatial weights (\(n \times  n\)). 

\subsection{Fisher's Z Transformation of Correlation}
Since our data are drawn from different samples from different periods, we apply the Fisher's Z Transformation of Correlation \cite{Fisher1915} to ensure that we can compare the chronological imbalance patterns for the proportion of tweets related to decentralized finance. 

The Fisher's Z Transformation of the Correlation coefficient is shown below\cite{welz2022fisher}:
\begin{equation}
    z = atanh(\rho) = 0.5ln{{1+\rho} \over {1-\rho}}
\end{equation}
Where \(\rho\) is the sample correlation coefficient. 

\subsection{Cluster Analysis}
Cluster analysis is a mathematical method to find out the similar characteristics in given sets and classify objects that have similar traits into one cluster while filtering out the dissimilar objects as possible as we can \cite{kaufman2009finding, romesburg2004cluster}. 

Among all the clustering methods, K-means clustering is one of the simplest methods to conduct clustering, as it focuses on defining k centroids, one for each cluster \cite{kodinariya2013review}. The K-means method intends to minimize an objective function, denoted as the following\cite{kodinariya2013review}. 

\begin{equation}
    W(S,C)=\sum_{k=1}^{K}\sum_{i \in S_{k}}\|y_{i}-c_{k}\|^2 
\end{equation}
Where \(S\) is a K-cluster partition of the entity set, \(y_{i}\) (\(i \in I\)) is the vectors of \(S\) in the M-dimension space, \(S_{k}\) is the non-empty non-overlapping clusters, \(c_{k}\) represents the centroid in each cluster,\( k=1,2,...K\). 

According to Kodinariya et al., \cite{kodinariya2013review}, K-means clustering requires the following steps. First, we need to put K points in our space, representing as initial group centroids, as far away from each other as possible. Second, assign each point to a given dataset and associate it with the nearest centroid. Third, after finishing assigning points, recalculate the new K centroids in the middle of the clusters. Next, repeat steps 2 and 3 until the centroids stop moving. 

In our research, we choose the Silhouette method to find the optimal number of clusters in K-means clustering. Silhouette Score, a method that computes the average of the Silhouette coefficient of all samples from different numbers of clusters, is the measurement of the clustering quality that ranges from -1 to 1, \cite{shahapure2020cluster}. Typically, when the average is close to 1, it means that all the points are in the correct cluster \cite{shahapure2020cluster}. Therefore, we should find the number of clusters that yield the highest Silhouette score. The formula of the Silhouette coefficient can be represented as the following.
\begin{equation}
    S=(b-a) \over max(a,b)
\end{equation}
Where \(b\) represents the mean nearest cluster distance for each point in our dataset, and \(a\) is the mean intra-cluster distance for each point.

\subsection{Topic Modeling}
Topic Modelling is a revolutionary method for text mining\cite{kherwa2019topic}. It is a powerful tool used in various fields to conduct text analysis to organize, summarize, and understand a gargantuan number of unorganized texts, images, audio, etc \cite{li2021bibliometric}. Common topic modeling methods encompass several techniques, including 1) Latent Semantic Analysis (LSA); 2) Non-Negative Matrix Factorization (NNMF); 3) Probabilistic Latent Semantic Analysis (PLSA); and 4) Latent Dirichlet Allocation (LDA) \cite{kherwa2019topic}.

In our article, we focus on the application of Latent Dirichlet Allocation (LDA), a classic and widely used method predating the advent of Large Language Models (LLMs). LDA is a probabilistic model that identifies underlying topics within a document by assuming that each document is composed of a mixture of topics, where each topic is characterized by a distribution over a vocabulary \cite{kherwa2019topic}. Assuming that we have M documents in a given set, and each document has \(N_{d}\) words, where \(d=1,2,...,M\), the following are the steps of conducting LDA \cite{jelodar2019latent}: 

\begin{enumerate}
    \item[\textbf{(1)}] Create a Dirichlet distribution \(\phi_{t}\) for topic \(t\) (\(t=1,2,...,T)\)) with parameter \(\beta\)
    \item[\textbf{(2)}] Create a Dirichlet distribution \(\sigma_{d}\) for document \(d\) with parameter \(\alpha\).
    \item[\textbf{(3)}] For each word \(W_{n}\) (\(n=1,2,...,N_{d})\)) in document \(d\)
\begin{enumerate}
    \item[\textbf{\textbf{a.}}] Select a topic \(Z_{n}\) from \(\sigma_{d}\)
    \item[\textbf{\textbf{b.}}] Select a word \(W_{n}\) from \(\phi_{zn}\)
\end{enumerate}
\end{enumerate}


\section{Results}
\label{sec:results}
The results revealed distinct patterns in the attention and sentiment related to decentralized finance on social media among various countries globally (RQ1). These patterns, correlating with economic factors represented by GDP per capita, uncovered a certain global imbalance. Additionally, tracking these changes from 2012 to 2022 has provided us with a deeper understanding of how this global imbalance evolves across time (RQ2). The topic modeling results indicate that tweets from countries of different clusters tend to revolve around relatively distinct topics when discussing decentralized finance, providing qualitative evidence to supplement our quantitative analysis (RQ3).

\begin{figure}[!htbp]
 \centering 
 \includegraphics[width=0.45\textwidth]{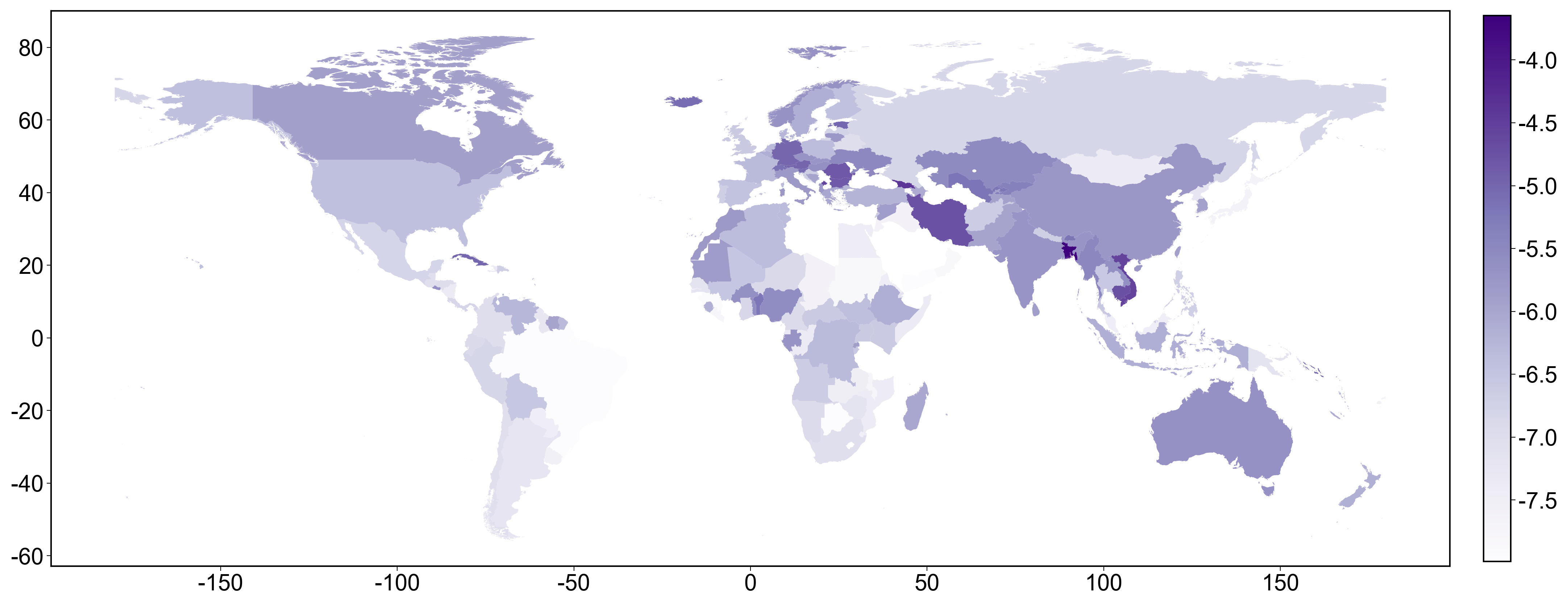}
  \caption{Proportions of tweets related to decentralized finance across countries. The figure illustrates the spatial distribution of tweets containing at least one decentralized finance-related keyword. Darker shades indicate a higher proportion of relevant tweets originating from that country, while lighter shades represent lower proportions.}
  \label{fig:map-proportion}
\end{figure}

\begin{figure}[!htbp]
\centering
  \includegraphics[width=0.45\textwidth]{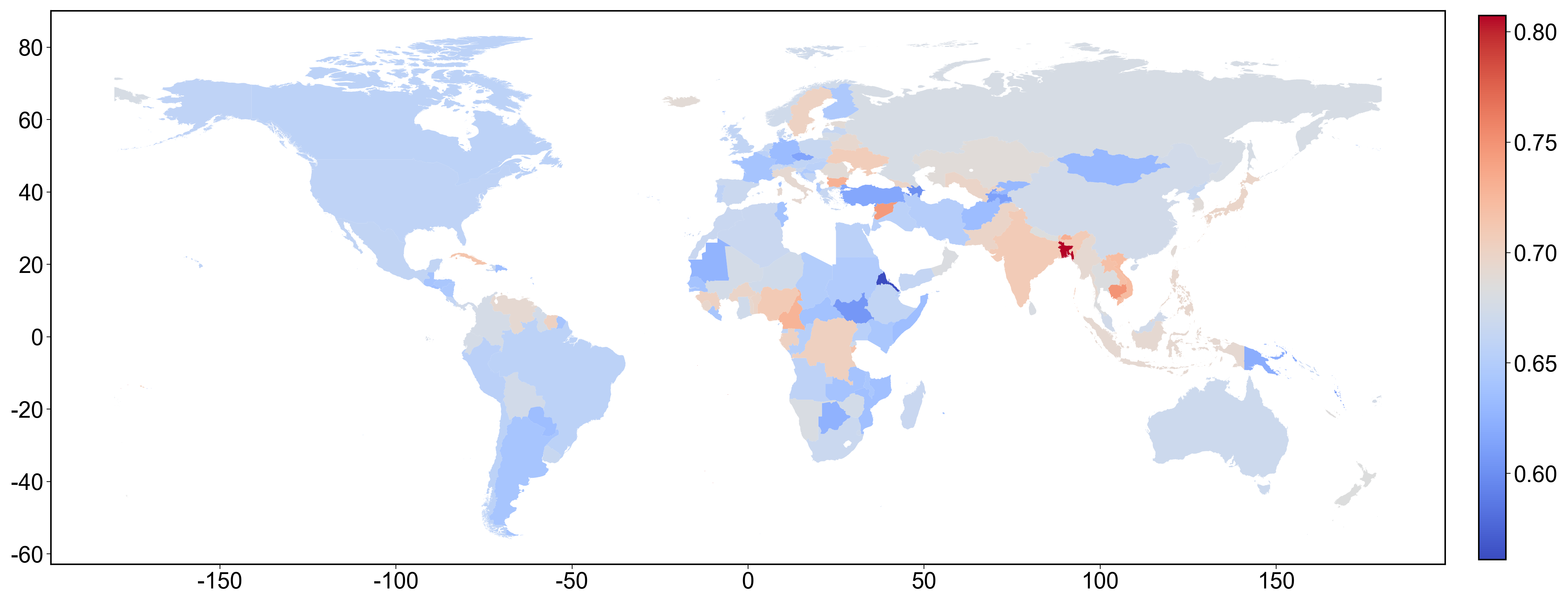}
  \caption{Sentiment scores of tweets related to decentralized finance across countries. The figure visualizes the average sentiment score of tweets containing at least one decentralized finance-related keyword. Higher sentiment scores (in darker shades) indicate a more positive sentiment, while lower scores (in lighter shades) reflect a more negative sentiment.}
  \label{fig:map-sentiment}
\end{figure}

\subsection{Spatial Autocorrelation}
Figures \ref{fig:map-proportion} and \ref{fig:map-sentiment} provide a geographic overview of decentralized finance discussions on Twitter. The first map highlights the proportion of relevant tweets per country, while the second visualizes their sentiment distribution. 

The results in Table \ref{tab:autocorrelation} elucidate that our variables and clusters exhibit spatial autocorrelation, as indicated by the relatively small p-values. This suggests that decentralized finance-related discussions are not randomly distributed across space but tend to cluster geographically. Consequently, a geographically weighted regression (GWR) is more appropriate than a simple linear regression for analyzing these patterns, as GWR accounts for spatial heterogeneity and localized variations in relationships between variables.

\begin{table*}[!htbp]
\centering
\caption{Spatial autocorrelation}
\label{tab:autocorrelation}
\begin{tblr}{
  width = \linewidth,
  colspec = {Q[184]Q[94]Q[130]Q[130]Q[130]Q[130]Q[130]},
  row{1} = {c},
  row{2} = {c},
  cell{1}{1} = {r=2}{},
  cell{1}{2} = {r=2}{},
  cell{1}{3} = {c=5}{0.69\linewidth},
  cell{3}{3} = {c},
  cell{3}{4} = {c},
  cell{3}{5} = {c},
  cell{3}{6} = {c},
  cell{3}{7} = {c},
  cell{4}{3} = {c},
  cell{4}{4} = {c},
  cell{4}{5} = {c},
  cell{4}{6} = {c},
  cell{4}{7} = {c},
  cell{5}{3} = {c},
  cell{5}{4} = {c},
  cell{5}{5} = {c},
  cell{5}{6} = {c},
  cell{5}{7} = {c},
  cell{6}{3} = {c},
  cell{6}{4} = {c},
  cell{6}{5} = {c},
  cell{6}{6} = {c},
  cell{6}{7} = {c},
  cell{7}{3} = {c},
  cell{7}{4} = {c},
  cell{7}{5} = {c},
  cell{7}{6} = {c},
  cell{7}{7} = {c},
  cell{8}{3} = {c},
  cell{8}{4} = {c},
  cell{8}{5} = {c},
  cell{8}{6} = {c},
  cell{8}{7} = {c},
  hline{1,3,9} = {-}{},
  hline{2} = {3-7}{},
}
Variable & Test & Weights &  &  &  & \\
 &  & 1-nearest neighbor & 2-nearest neighbor & 3-nearest neighbor & 4-nearest neighbor & 5-nearest neighbor\\
ln(GDP per capita) & Moran's I & 0.800*** (0.100) & 0.751*** (0.075) & 0.704*** (0.061) & 0.687*** (0.054) & 0.675*** (0.047)\\
ln(Tweet proportion) & Moran's I & 0.357*** (0.101) & 0.368*** (0.074) & 0.358*** (0.061) & 0.340*** (0.053) & 0.328*** (0.047)\\
Sentiment score & Moran's I & 0.282** (0.100) & 0.162* (0.073) & 0.151** (0.059) & 0.156** (0.051) & 0.160*** (0.047)\\
Cluster 1 (binary) & Chi-square & 4.700* & 7.298** & 10.980*** & 12.285*** & 12.159***\\
Cluster 2 (binary) & Chi-square & 5.686* & 15.614*** & 19.548*** & 25.459*** & 27.293***\\
Cluster 3 (binary) & Chi-square & 22.697*** & 38.920*** & 46.047*** & 66.707*** & 72.755***
\end{tblr}

\raggedright
\small \textit{Note}: This table presents the results of the spatial autocorrelation analysis using Moran’s I and Chi-square tests across different spatial weight structures. \textit{Moran’s I for GDP per capita, tweet proportion, and sentiment score}: The positive and statistically significant Moran’s I values indicate strong spatial clustering, meaning that similar values of these variables tend to be geographically concentrated. The effect is strongest for GDP per capita, suggesting that wealthier countries tend to cluster together. The spatial autocorrelation for tweet proportion and sentiment score, while weaker, remains significant across different nearest-neighbor specifications. \textit{Chi-square tests for country clusters}: The spatial distribution of country clusters is tested using Chi-square statistics. Higher values indicate stronger clustering patterns. Cluster 3 (high-GDP countries) exhibits the highest level of spatial clustering, followed by Cluster 2 and then Cluster 1. This suggests that economic and social factors influencing decentralized finance discussions are not randomly distributed but form distinct spatial patterns. \textit{Nearest-neighbor approach}: The table reports results using 1 to 5 nearest neighbors to define spatial relationships. The decreasing magnitude of Moran’s I values as more distant neighbors are included suggests that spatial autocorrelation weakens with distance. \textit{Statistical significance levels}: $p$ $<$ 0.05 (*), $p$ $<$ 0.01 (**), and $p$ $<$ 0.001 (***) indicate the robustness of these spatial effects.
These findings emphasize that economic and social characteristics influencing decentralized finance discussions exhibit strong spatial dependencies. This reinforces the need for geographically weighted models when analyzing these trends.

\end{table*}

\subsection{Geographically Weighted Regression}
\label{sec: Geographically Weighted Regression}
We conclude that GDP per capita can significantly predict the proportion of tweets related to decentralized finance across countries after 2014.
From Table \ref{tab:regression}, based on the regression results, from 2012 to 2014, GDP per capita is not statistically significant with a relatively high p-value. However, from 2014 to 2022, GDP becomes a significant variable that affects the proportion of tweets related to decentralized finance across countries. Not surprisingly, after 2015 (including 2015), the GDP per capita is positively correlated to the proportion of tweets related to DeFI, indicating that the countries that have higher GDPs discuss more decentralized finance in tweets. For example, in 2022, a one percent increase in GDP will yield a 0.244 percent increase in the proportion of tweets related to DeFi. The quantitative impact of GDP per capita rose yearly from 2015 to 2017 and reached its highest in 2017. After that, the impact displays a decreasing trend. In 2017, a one per cent increment in GDP per capita resulted in a 0.396 percent rise in the proportion, but in 2022, it only generated a 0.244 percent increase.

\begin{table*}[!htbp]
\centering
\caption{SEM regressions for proportion of tweets related to decentralized finance across countries for each year}
\label{tab:regression}
\begin{tabular}{>{\hspace{0pt}}m{0.15\linewidth}>{\hspace{0pt}}m{0.08\linewidth}>{\hspace{0pt}}m{0.08\linewidth}>{\hspace{0pt}}m{0.08\linewidth}>{\hspace{0pt}}m{0.08\linewidth}>{\hspace{0pt}}m{0.08\linewidth}>{\hspace{0pt}}m{0.08\linewidth}} 
\hline
Year & 2012 & 2013 & 2014 & 2015 & 2016 & 2017 \\ 
\hline
Constant & 0.006 & 0.003 & 0.003 & 0.014 & 0.006 & 0.006 \\ 
ln(GDP per capita) & -0.183 (0.111) & 0.168 (0.086) & 0.033 (0.093) & 0.281** (0.092) & 0.387*** (0.085) & 0.396*** (0.089) \\ 
lambda & -0.074 & 0.056 & 0.162 & 0.236 & 0.104 & 0.186 \\ 
Pseudo $R^{2}$ & 0.036 & 0.030 & 0.001 & 0.085 & 0.156 & 0.154 \\ 
N & 71 & 141 & 144 & 143 & 136 & 138 \\ 
\hline
Year & 2018 & 2019 & 2020 & 2021 & 2022 &  \\ 
\hline
Constant & 0.006 & 0.010 & -0.005 & -0.001 & 0.009 &  \\ 
ln(GDP per capita) & 0.371*** (0.089) & 0.235* (0.095) & 0.207* (0.094) & 0.274** (0.097) & 0.244* (0.097) &  \\ 
lambda & 0.217 & 0.373 & 0.274 & 0.293 & 0.400 &  \\ 
Pseudo $R^{2}$ & 0.146 & 0.098 & 0.088 & 0.084 & 0.114 &  \\ 
N & 142 & 142 & 141 & 136 & 133 &  \\
\hline
\end{tabular}

\raggedright
\small \textit{Note}: Standard errors are presented in parentheses. \textit{*} $p$ $<$ 0.05, \textit{**} $p$ $<$ 0.01, \textit{***} $p$ $<$ 0.001. \textit{GDP per capita} represents the economic output per person for each country in each year, measured in logarithmic form to adjust for skewness in income distribution. \textit{Tweet proportion} represents the volume of tweets related to decentralized finance, normalized by the total volume of tweets in each country for each year. \textit{Spatial Lag ($\lambda$)} measures the degree of spatial dependence in the regression model. A positive $\lambda$ suggests that decentralized finance tweet proportions in a given country are influenced by neighboring countries’ tweet proportions, indicating regional clustering effects. \textit{Pseudo $R^{2}$} is used to assess the model’s explanatory power. The increase in $R^{2}$ over the years suggests a growing relationship between GDP per capita and decentralized finance discussions.\textit{Key Observations}: Before 2015, GDP per capita had an inconsistent effect on the proportion of decentralized finance-related tweets, with some years showing negative or insignificant relationships; Starting in 2015, the effect of GDP per capita became statistically significant, indicating that economic development increasingly influenced decentralized finance discussions; The highest coefficients occur between 2016 and 2018, suggesting a peak period where wealthier countries saw greater public engagement in decentralized finance discussions on Twitter; The spatial lag coefficient ($\lambda$) grows stronger in later years, suggesting that regional spillover effects in decentralized finance discussions became more pronounced. These findings highlight the evolving relationship between economic development and decentralized finance discussions and emphasize the role of spatial dependencies in shaping global Twitter activity trends.
\end{table*}

\begin{figure}[!htbp]
\centering
\includegraphics[width=0.45\textwidth]{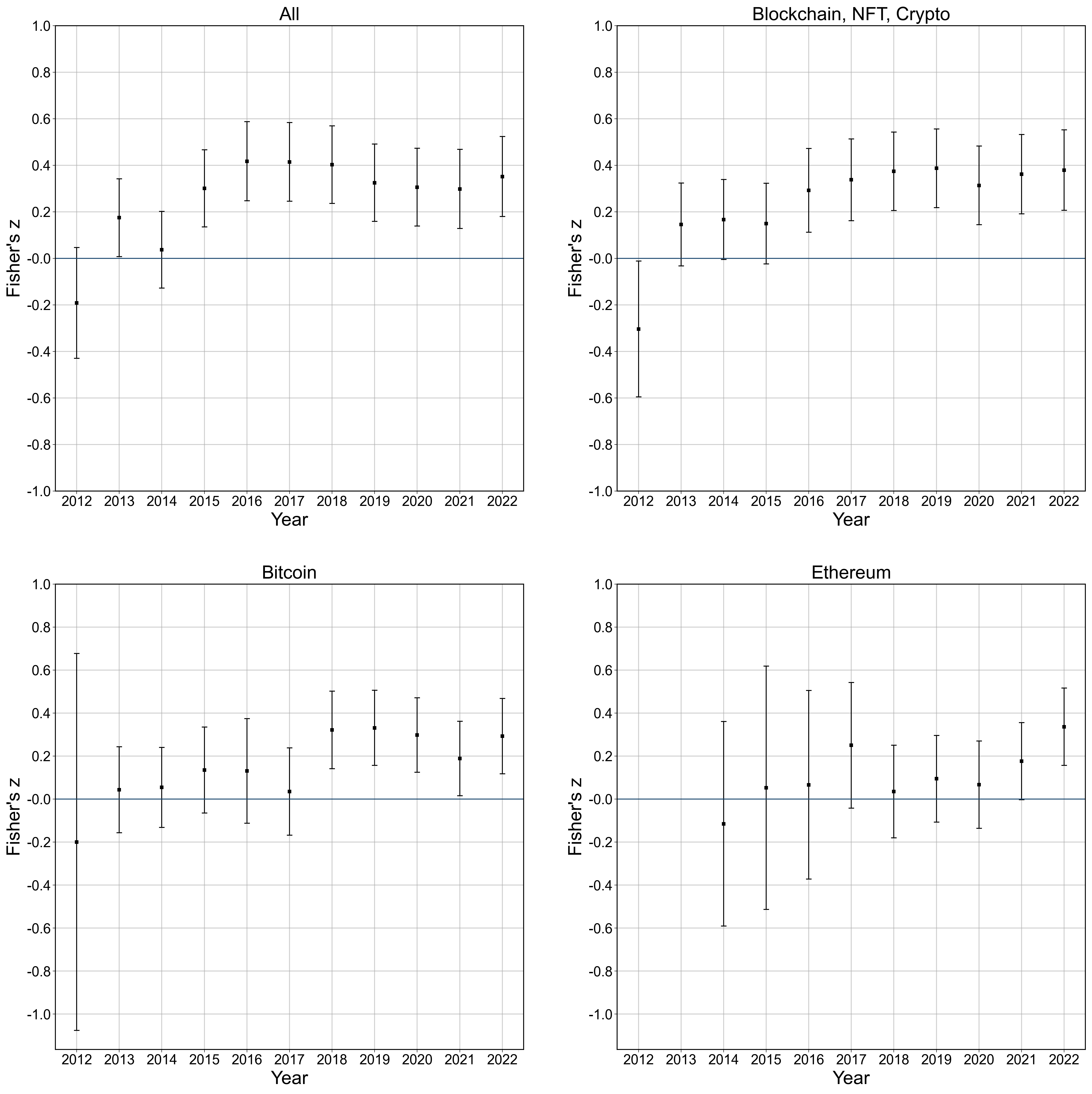}
  \caption{Fisher's transformation of correlation coefficients across years. The figure illustrates the temporal evolution of the correlation between GDP per capita and the proportion of tweets containing keywords related to decentralized finance. The Fisher Z-transformed correlation coefficients provide a stabilized measure of correlation, accounting for variations in sample size across years. Positive values indicate a stronger association between GDP per capita and keyword prevalence, while negative values suggest an inverse relationship. The confidence intervals represent the statistical uncertainty in estimating these correlations over time.}
  \label{fig:fisher}
\end{figure}

\subsection{Fisher's Z Transformation of Correlation}
Figure \ref{fig:fisher} presents the results of our robustness checks, which examine the relationship between GDP per capita and the proportion of tweets related to decentralized finance keywords over time. Instead of analyzing the total volume of related tweets, we focus on specific keyword groups to assess whether economic factors influence discussions differently depending on the terminology used.

The results indicate that when considering general keywords such as \textit{blockchain}, \textit{NFT}, and \textit{crypto}, GDP per capita maintains a positive and statistically significant correlation with the proportion of tweets containing these keywords, particularly after 2015. This suggests that higher-income countries tend to have a greater prevalence of general decentralized finance discussions on social media.

However, when narrowing the analysis to more specific terms such as \textit{Ethereum}, GDP per capita does not exhibit a statistically significant effect on the proportion of tweets. This discrepancy suggests that while economic development may influence general awareness and engagement with decentralized finance, discussions around specific projects or technologies may be driven by other factors, such as niche communities, technological adoption, or speculative interest.


\subsection{Spatial Clustering}
\label{sec: Clustering}

\begin{figure}[!htbp]
\centering
  \includegraphics[width=0.45\textwidth]{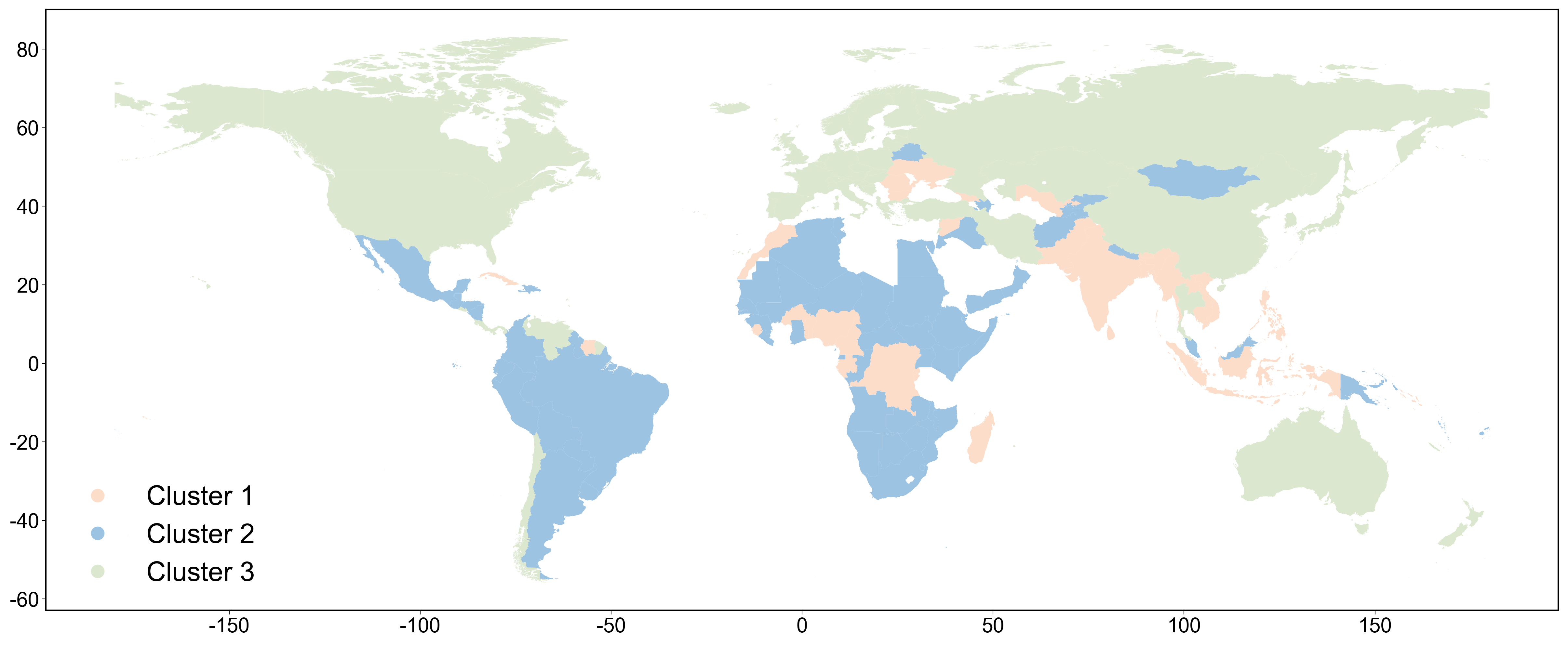}
  \caption{Global clustering of countries based on decentralized finance-related tweets, sentiment scores, and economic indicators. Countries are color-coded according to their assigned cluster, which reflects similarities in their level of engagement with decentralized finance, public sentiment, and economic development. The classification is derived using a K-means clustering algorithm on 145 countries.}
  \label{fig:map-clustering}
\end{figure}

\begin{figure}[!htbp]
\centering
  \includegraphics[width=0.45\textwidth]{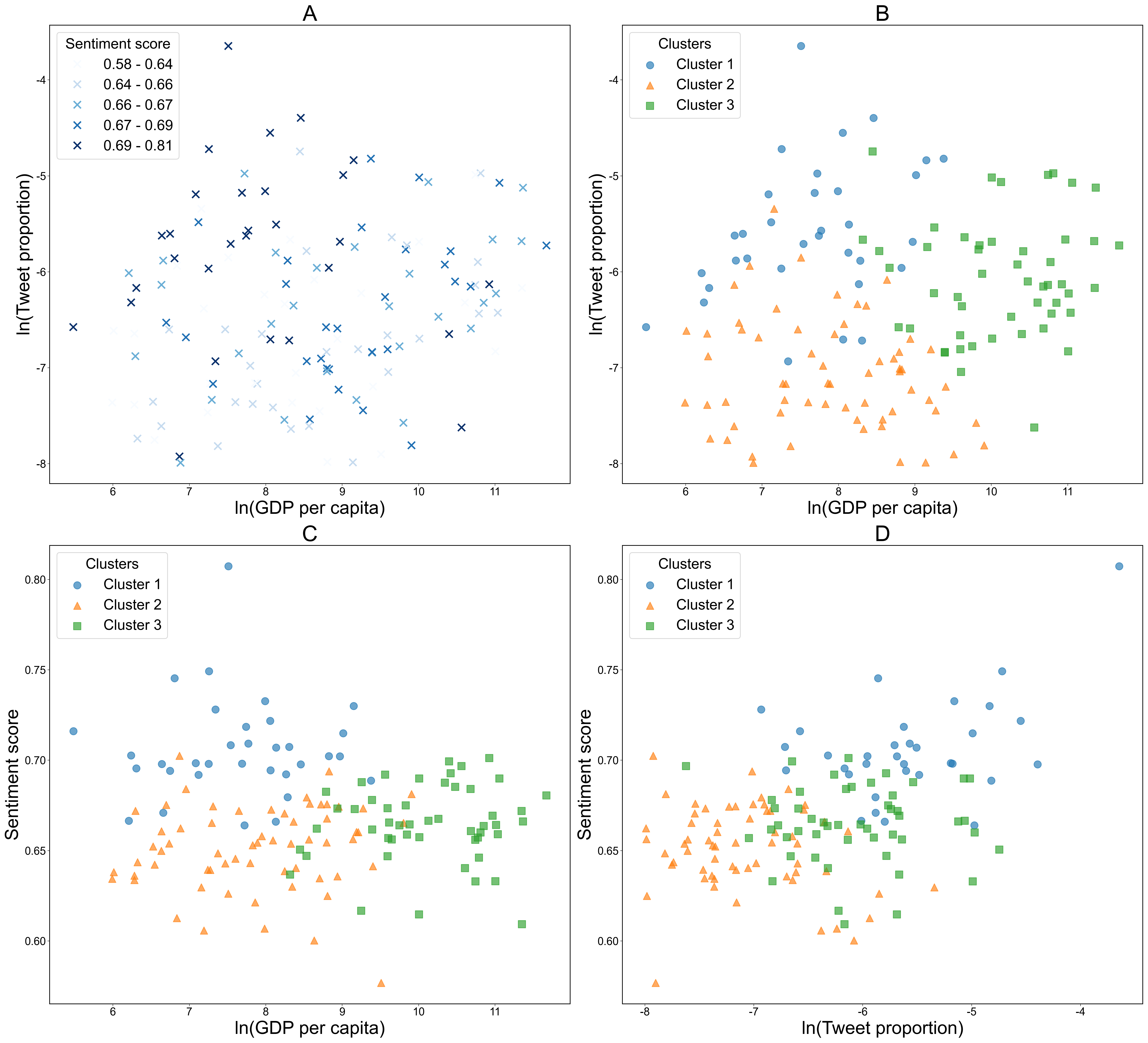}
  \caption{Scatter plot of GDP per capita, sentiment score, and proportion of tweets related to decentralized finance. Each point represents a country, positioned according to its GDP per capita and the proportion of relevant tweets, with color intensity reflecting sentiment score. The visualization highlights patterns in economic development, online engagement with decentralized finance, and public sentiment across different countries.}
  \label{fig:scatter}
\end{figure}

\begin{figure*}[!htbp]
\centering
  \includegraphics[width=\textwidth]{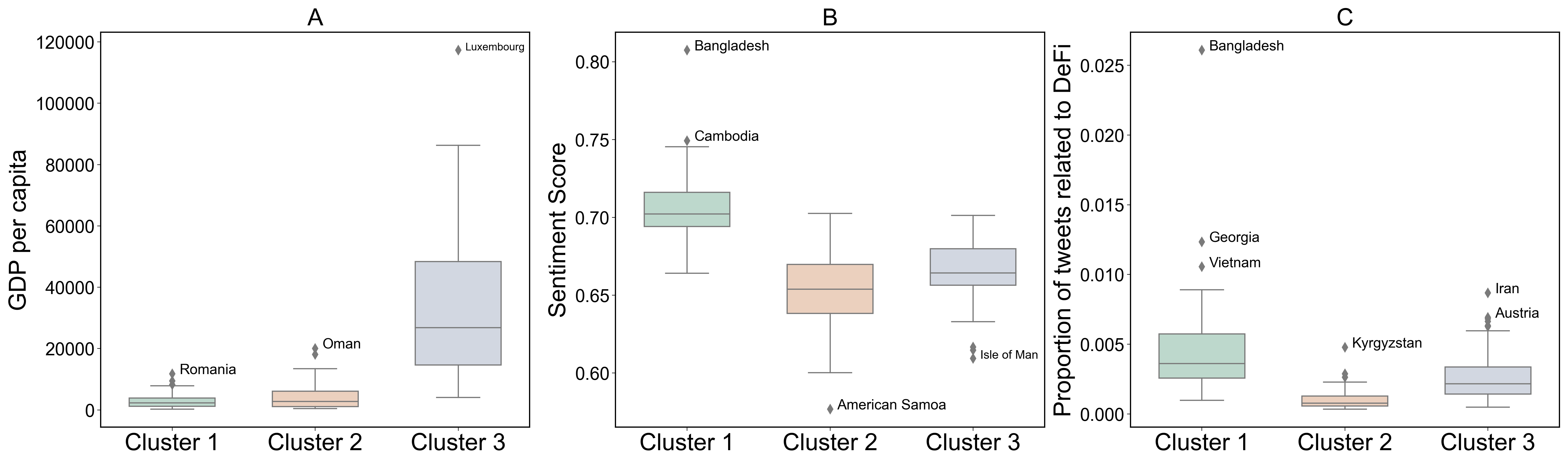}
  \caption{Box plot of GDP per capita, sentiment score, and proportion of tweets related to decentralized finance for each country cluster. The figure provides a statistical summary of the three key variables used in the clustering analysis, allowing for comparison between the three identified country clusters. The distribution of GDP per capita shows significant variation across clusters, with Cluster 3 containing the highest-income countries. The sentiment scores and tweet proportions also exhibit distinct trends, illustrating differences in engagement with decentralized finance across economic levels.}
  \label{fig:clusters_box}
\end{figure*}

\begin{table}[!h]
\centering
\caption{Descriptive Statistics for Different Clusters}
\label{tab:clustering}
\renewcommand{\arraystretch}{1.2} 
\setlength{\tabcolsep}{8pt} 
\begin{tabular}{lccc} 
\hline
\textbf{Variable} & \textbf{Cluster 1} & \textbf{Cluster 2} & \textbf{Cluster 3} \\ 
\hline
\multicolumn{4}{l}{\textbf{GDP per Capita (USD)}} \\ 
\hline
Mean & 3,075 & 4,222 & 33,879 \\ 
Median & 2,256 & 2,748 & 26,785 \\ 
Min & 240 & 399 & 4,099 \\ 
Max & 11,795 & 20,064 & 117,302 \\ 
Std Dev & 2,811 & 4,203 & 25,430 \\ 
\hline
\multicolumn{4}{l}{\textbf{Tweet Proportion (\%)}} \\ 
\hline
Mean & 0.497 & 0.102 & 0.280 \\ 
Median & 0.361 & 0.077 & 0.217 \\ 
Min & 0.098 & 0.034 & 0.049 \\ 
Max & 2.608 & 0.478 & 0.869 \\ 
Std Dev & 0.470 & 0.075 & 0.186 \\ 
\hline
\multicolumn{4}{l}{\textbf{Sentiment Score}} \\ 
\hline
Mean & 0.706 & 0.651 & 0.665 \\ 
Median & 0.702 & 0.654 & 0.664 \\ 
Min & 0.664 & 0.577 & 0.609 \\ 
Max & 0.807 & 0.702 & 0.701 \\ 
\hline
\textbf{Number of Countries} & 33 & 62 & 50 \\ 
\hline
\end{tabular}
\\
\raggedright
\small \textit{Note}: GDP per capita represents the average value for each country from 2012 to 2022. The tweet proportion is determined by the volume of tweets related to decentralized finance, normalized by the total volume of tweets in each country from 2012 to 2022. The sentiment score is the average value of tweets related to decentralized finance for each country from 2012 to 2022. 
\end{table}

We found out that classifying countries into three clusters yielded the highest Silhouette Score of 0.312. Consequently, we conducted a K-means cluster analysis on 145 countries, categorizing them into three distinct clusters based on the proportion of tweets related to decentralized finance, the average sentiment score of those tweets, and GDP per capita as the primary features. Given that both GDP per capita and tweet proportion exhibit long-tail distributions, we applied a logarithmic transformation to these variables prior to further analysis to better approximate a normal distribution. The Silhouette scores for different numbers of clusters are shown in Figure \ref{fig:silhouette} in the Appendix. A list of the countries included in each cluster is provided in Appendix \ref{sec: Clusters of Countries}. Figure \ref{fig:map-clustering} illustrates the geographical distribution of countries belonging to different clusters. Table \ref{tab:clustering} presents the descriptive statistics for these three clusters. 

\subsubsection{Descriptive Statistics}
\label{descriptive statistics}
Countries in Cluster 1 and Cluster 2 are characterized by relatively underdeveloped economies, with average GDP per capita values of \$3,075 and \$4,222, respectively. Our regression results in Table \ref{tab:regression} indicate that countries with more advanced economic development tend to have higher public attention towards decentralized finance. However, countries in Cluster 1 present an interesting anomaly, exhibiting the highest level of attention to decentralized finance on social media, as reflected by the average proportion of related tweets (0.497\%) and the most positive sentiment (average sentiment score of 0.706). In contrast, countries in Cluster 2 demonstrate the lowest level of attention to decentralized finance (average proportion of related tweets of 0.102\%) and a relatively low sentiment (average sentiment score of 0.651).

Countries in Cluster 3, which have the most developed economies (average GDP per capita of \$33,879), show levels of attention to decentralized finance (average proportion of related tweets of 0.280\%) and sentiment scores (average sentiment score of 0.665) that are lower than those in Cluster 1 but higher than those in Cluster 2. Figure \ref{fig:clusters_box} presents box plots depicting various countries based on the three variables mentioned above. 

Cluster 1 reveals the lowest GDP situation, representing the least economically developed countries and exhibiting the lowest median GDP. Countries in Cluster 2 have slightly higher GDP levels than those in Cluster 1, as indicated by a higher median and elevated position of each percentile. The GDP levels of countries in Cluster 3 surpass those of both Cluster 1 and Cluster 2 significantly. Based on these GDP disparities, we classify Cluster 1 as the lowest-GDP countries, Cluster 2 as the mid-GDP countries, and Cluster 3 as the highest-GDP countries.

\subsubsection{Crypto Legal Status}
\label{sec: Crypto Legal Status}

\begin{figure}[!htbp]
\centering
\includegraphics[width=0.49\textwidth]{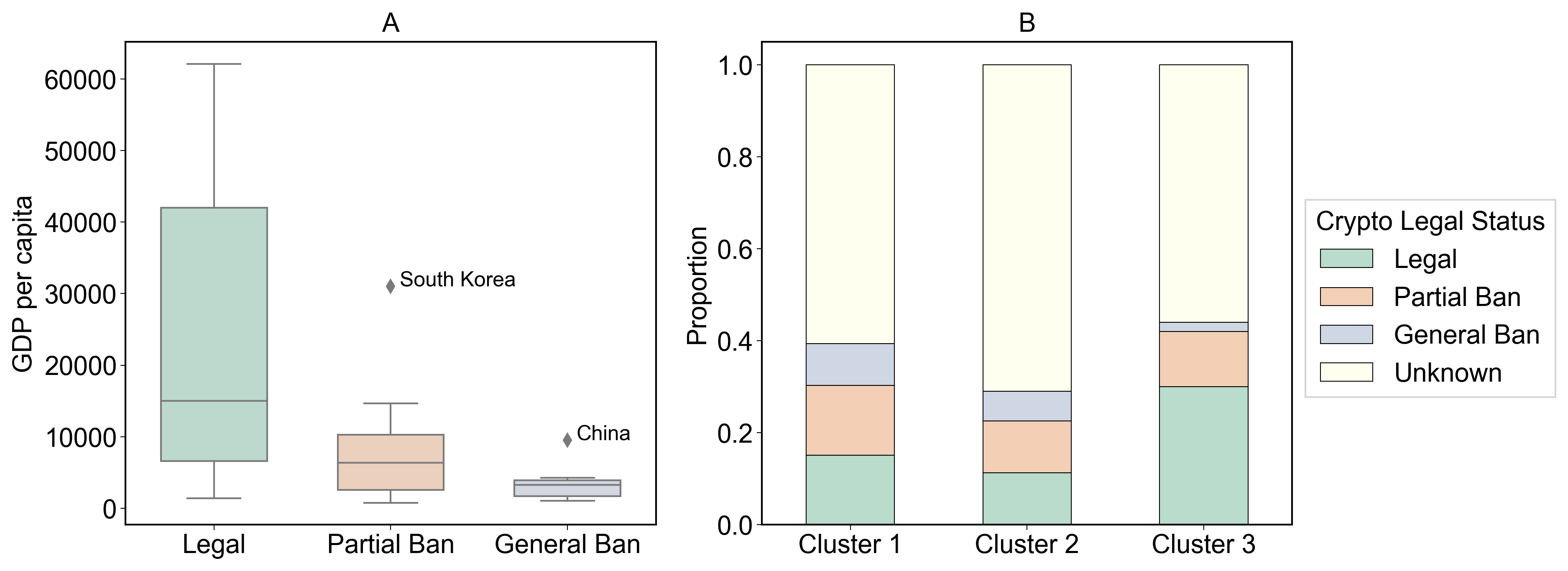}
  \caption{Cryptocurrency legal status across clusters. (A) Box plots illustrate the relationship between GDP per capita and the legal status of cryptocurrency trading in different countries. Countries that fully recognize cryptocurrencies tend to have the highest GDP per capita, while those with a partial or general ban exhibit decreasing economic development. (B) The bar chart presents the proportion of countries in each cluster that recognize the legality of cryptocurrencies. The most economically developed cluster (Cluster 3) has the highest percentage of countries where cryptocurrency trading is legal, whereas Clusters 1 and 2, with lower GDP per capita, exhibit greater regulatory restrictions.}
  \label{fig:legal-status}
\end{figure}

Considering the varying legal statuses of cryptocurrency trading in different countries, which may influence the level of attention and sentiment towards topics related to decentralized finance, we incorporated the "Crypto Legal Status" data from the "Cryptocurrency Regulation Tracker". Box plots (Figure \ref{fig:legal-status}A) reveal a correlation between GDP per capita and the legality of cryptocurrencies. Excluding the outliers, South Korea and China, countries that recognize the legality of cryptocurrencies, those with a partial ban, and those with a general ban on cryptocurrencies show decreasing levels of economic development, with average GDP per capita values of \$24,009, \$7,635, and \$3,565, respectively. 
We assigned values to the three crypto legal statuses (Legal, Partial Ban, General Ban) as 0, 0.5, and 1, respectively, to represent the degree of prohibition of cryptocurrencies. The Spearman correlation coefficient between this variable and GDP per capita is -0.530 (p-value $<$ 0.001). 
The proportions of countries within different clusters that recognize the legality of cryptocurrency (Figure \ref{fig:legal-status}B) also show a similar pattern. In the most economically developed Cluster 3, 30\% (15 out of 50) of the countries recognize the legality of cryptocurrency trading. In contrast, only 15\% (5 out of 33) in Cluster 1 and 11\% (7 out of 62) in Cluster 2 do so.



\begin{figure*}[!htbp]
\centering
\includegraphics[width=0.65\textwidth]{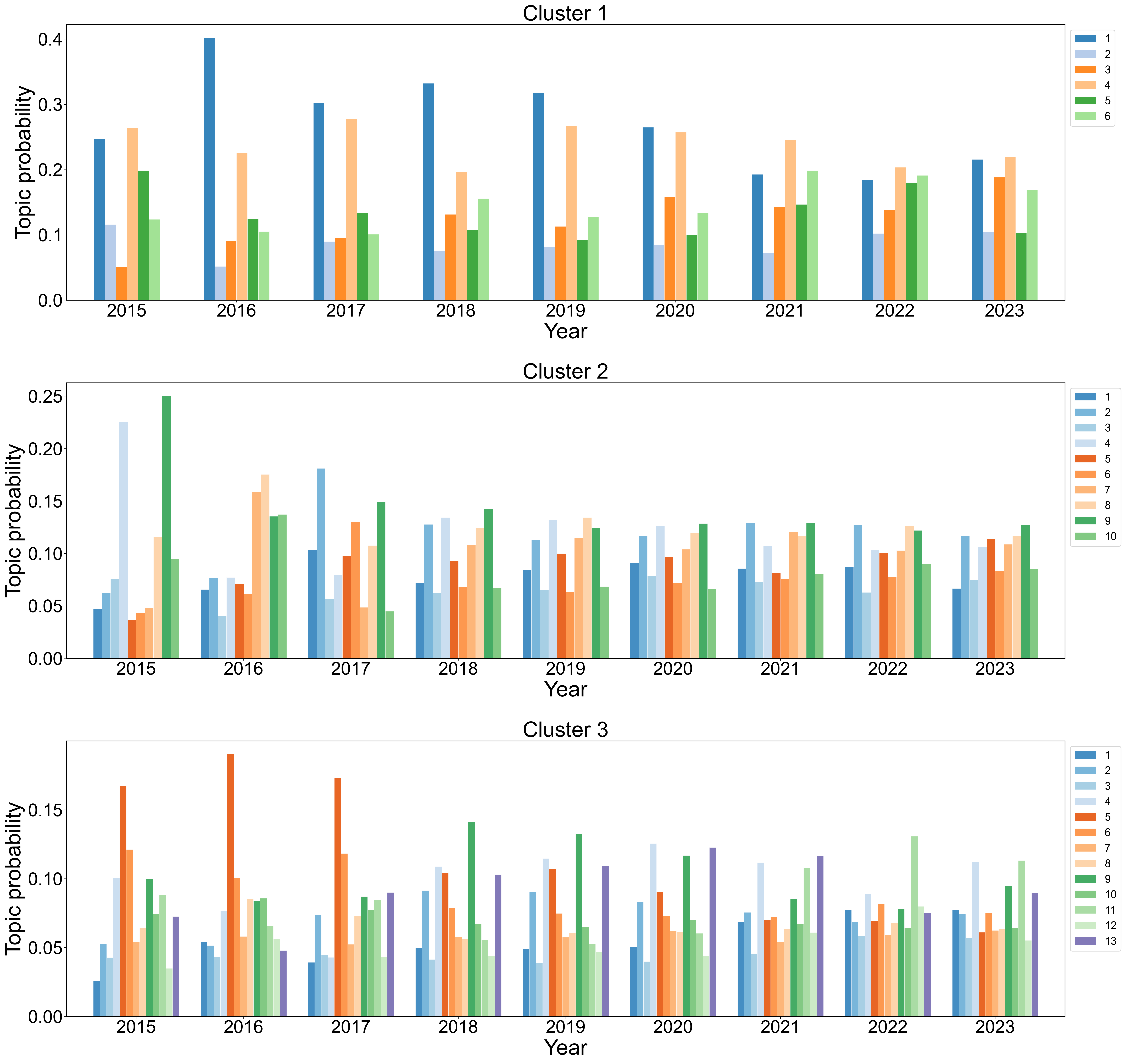}
  \caption{Temporal evolution of topic probabilities for each cluster. This figure illustrates how discussions around decentralized finance have evolved across different economic clusters over the years. The topic probability trends reveal distinct thematic preferences among countries with different economic backgrounds. Countries in Cluster 1, despite their lower GDP per capita, exhibit sustained interest in specific aspects of decentralized finance, while mid- and high-GDP clusters display more varied topic distributions over time.}
  \label{fig:topic-modeling}
\end{figure*}

\begin{figure*}[!htbp]
\centering
\includegraphics[width=0.9\textwidth]{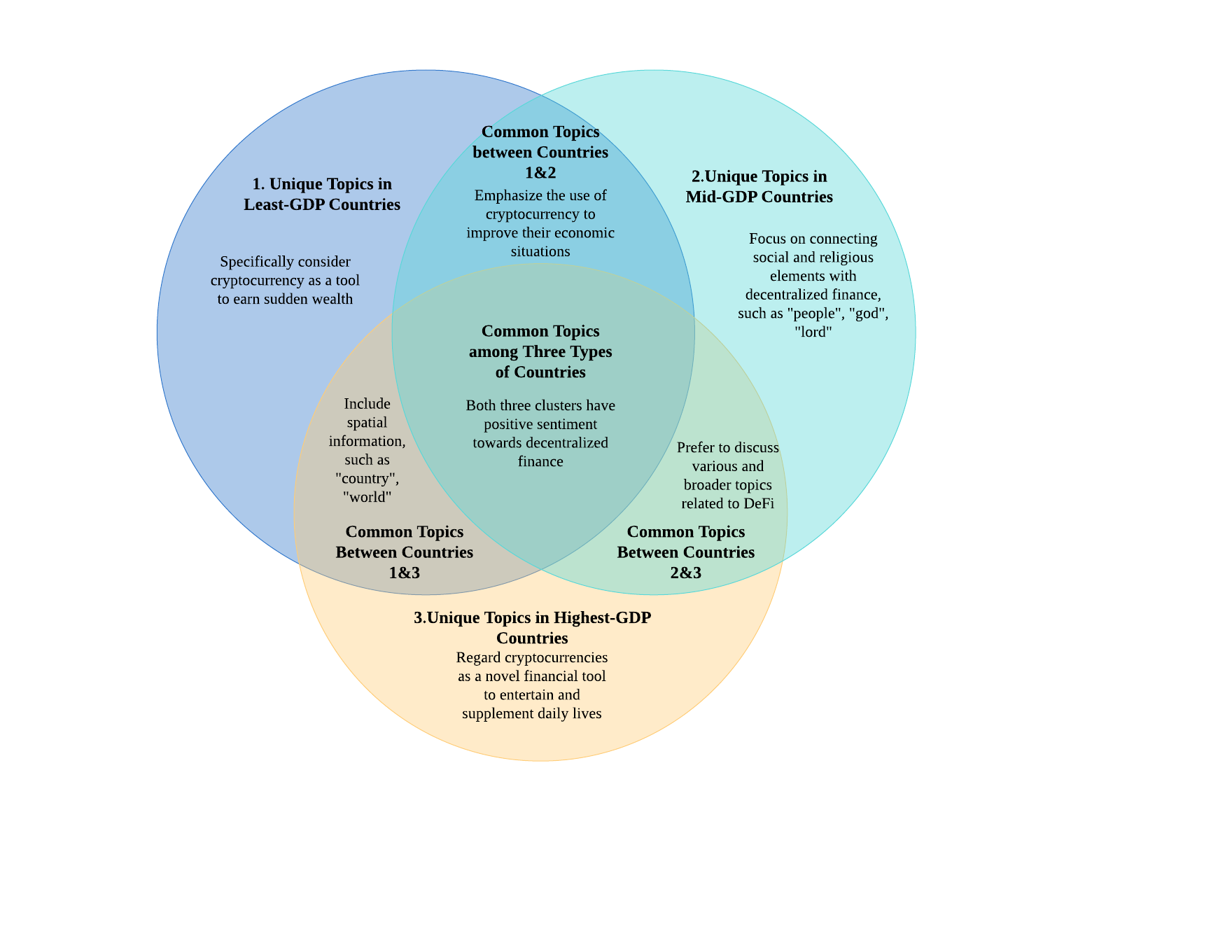}
  \caption{Comparison of shared and unique topics in decentralized finance discussions across clusters. This figure presents a Venn diagram-style analysis of the most frequently discussed topics within each cluster. While all clusters share a core interest in decentralized finance, their discussions diverge significantly in focus. Lowest-GDP countries (Cluster 1) emphasize financial hope and speculative gains, mid-GDP countries (Cluster 2) associate decentralized finance with social and religious aspects, and highest-GDP countries (Cluster 3) discuss cryptocurrencies primarily as investment and entertainment tools. Despite these distinctions, common themes such as financial innovation and blockchain adoption remain prevalent across all groups.}
  \label{fig:topic-modeling-venn}
\end{figure*}

The results from Figure \ref{fig:topic-modeling} reveal that countries in different clusters prefer different topics over time. The optimal number of topics for each cluster is based on coherence scores (Figure \ref{fig:coherence} in Appendix \ref{appendix:topic_modelling}). Based on the table of top words for different topics in different clusters in Appendix \ref{appendix:topic_modelling}, we summarized the differences and similarities in the top words among the three types of clusters.

Although the countries in all three clusters generally have a positive sentiment towards decentralized finance, the countries in Cluster 1 engage in more focused discussions and more frequently use words with positive emotions. This is consistent with our findings in Section \ref{sec: Clustering}: while GDP per capita is positively related to public attention, the countries in Cluster 1 show erratic public attention patterns, revealing the highest public attention and lowest average GDP per capita simultaneously.

Both lowest-GDP and mid-GDP countries emphasize the use of cryptocurrency to improve their economic situations, but the lowest-GDP countries surprisingly may consider cryptocurrency as a tool to achieve sudden wealth, characterized by words like "luck", "hope", and "promote". The mid-GDP countries focus on connecting social and religious elements with decentralized finance, using terms such as "people", "god", and "lord". However, the highest-GDP countries regard cryptocurrencies as novel financial tools to entertain and supplement daily life.

Despite focusing on different topics related to decentralized finance, the three types of countries share some similarities (see Figure~\ref{fig:topic-modeling-venn}).


\section{Discussion}
\label{sec:discussion}
\subsection{Limitations}
Regarding our DeFi dataset obtained from Harvard University's Twitter Archive v2.0, we identified a potential bias arising from the disparity in the population of Twitter users across different countries. This disparity could skew our results by introducing biased sentiment scores and an inaccurate proportion of tweets related to DeFi. We suspect that in less developed countries, a smaller fraction of the population uses Twitter as a daily social media platform for communication and idea exchange, which could lead to a lower proportion of discussions related to DeFi. For example, in the United States, regions with high Twitter usage tend to be concentrated in large cities with higher population densities. The uneven distribution of Twitter users is influenced not only by population density but also by factors such as gender and age \cite{jiang2019understanding}.

To further investigate this potential bias, we conducted a Pearson correlation analysis (see Fig. \ref{map-user} and Fig. \ref{linear-user} in Appendix F) between the total number of Twitter users and those posting DeFi-related tweets. The results indicate a positive correlation, suggesting that the bias in our DeFi-related tweets dataset is consistent with the inherent bias of Twitter itself. However, the demographic bias in Twitter usage is widely recognized and unavoidable \cite{jiang2019understanding}. While geotagged users may not represent a random sample of the population within a specific block group, the detailed and varied demographic data available allow for a reasonable approximation of the demographics of geotagged users in that area \cite{malik2015population}.

We compared the statistics of our two outliers, China and South Korea, using box plots from Figure \ref{fig:legal-status} alongside different clusters (see Figure \ref{fig:comparison-table}). Our geotagged tweets data may face challenges in accurately predicting sentiment and public attention in China due to the ban on Twitter within the mainland. Although some individuals access Twitter through VPNs (Virtual Private Networks), the majority of Chinese citizens have limited knowledge of VPN usage. Consequently, the sentiment and public attention data for China derived from geotagged tweets may be biased and inaccurate, reflecting only the viewpoints of a small segment of the population regarding decentralized finance.

\begin{table*}[!htbp]
\centering
\caption{Comparison of Statistics Between China, South Korea, and Different Clusters}
\label{fig:comparison-table}
\renewcommand{\arraystretch}{1.2} 
\setlength{\tabcolsep}{10pt} 
\begin{tabular}{lccc}
\toprule
\textbf{Country/Cluster} & \textbf{GDP per Capita (USD)} & \textbf{Tweet Proportion (\%)} & \textbf{Sentiment Score} \\ 
\midrule
China           & 9,538  & 0.003  & 0.673  \\ 
South Korea     & 31,038 & 0.003  & 0.688  \\ 
\midrule
Cluster 1 (mean) & 3,075  & 0.497  & 0.706  \\ 
Cluster 2 (mean) & 4,222  & 0.102  & 0.651  \\ 
Cluster 3 (mean) & 33,879 & 0.280  & 0.665  \\ 
\bottomrule
\end{tabular}

\raggedright
\small \textit{Note}: This table compares the GDP per capita, tweet proportion, and sentiment score of China and South Korea with the mean values of the three identified clusters. China and South Korea were identified as outliers in our analysis, as visualized in Figure \ref{fig:legal-status} and further compared using box plots in Figure \ref{fig:comparison-table}. China's geotagged tweet data may not accurately reflect public sentiment and attention due to the country's Twitter ban. While some individuals access Twitter through VPNs (Virtual Private Networks), the vast majority of Chinese citizens have limited knowledge of or access to VPN services. Consequently, the sentiment and public attention data for China likely represent a small, tech-savvy, and potentially non-representative subset of the population, which may introduce bias in interpreting decentralized finance discussions. South Korea, on the other hand, has unrestricted Twitter access, making its tweet data more reflective of general public opinion. However, despite its high GDP per capita, its tweet proportion remains significantly lower than those in the highest-income cluster (Cluster 3). This suggests that factors beyond economic development, such as cultural or regulatory influences, may shape decentralized finance engagement in South Korea. Overall, the comparison highlights the complexities in using geotagged tweets as a proxy for decentralized finance discussions, particularly in countries with restricted social media environments.
\end{table*}

\subsection{Implications}
\subsubsection{Poverty Reduction}

While our analysis reveals that economically advanced countries are more likely to engage in discussions about DeFi technologies, the rise and development of cryptocurrency suggest its potential to contribute to poverty reduction through unique and innovative features. DeFi technologies could serve as effective tools in mitigating the global economic and developmental imbalances highlighted by our findings on public attention and sentiment.

First, the transparency inherent in blockchain technology has the potential to reduce corruption in the public sector, particularly in developing countries. Corruption is widely recognized as a significant factor that impoverishes developing nations by exacerbating poverty, hindering economic growth, and impeding democratic transitions \cite{aina2014corruption, n2005causality, chetwynd2003corruption}. Blockchain, which underpins cryptocurrency, operates without central intermediaries and ensures that all transactions are transparently recorded in an immutable digital ledger. This system reduces the likelihood of manipulation, interference, or corruption by individuals, companies, or governments. Consequently, transactions occur directly between individuals, and the transparent environment helps to reduce obscure transactions and potential corruption in public sectors.

Second, cryptocurrency can act as a safeguard against the devaluation of wealth caused by hyperinflation. Like corruption, hyperinflation is another key factor that exacerbates poverty \cite{larochelle2014inter, ferreira1985poverty}. When governments print excessive amounts of money, they depreciate their currencies and drive up the prices of goods and services, thereby eroding purchasing power and deepening poverty. Cryptocurrencies offer a potential solution to hyperinflation by serving as a store of value. Their decentralized nature prevents manipulation by monetary authorities, ensuring that they are not subject to the effects of traditional monetary policies. Moreover, cryptocurrencies are more efficient than fiat currencies, enabling faster transactions and easier access. Their unique characteristics make them inherently resistant to hyperinflation \cite{lu2022cryptocurrency}.

\subsubsection{Sustainability}

The increasing attention and positive sentiment toward DeFi technologies in economically advanced regions underscore their growing importance and popularity. However, the widespread adoption and use of DeFi, particularly in crypto mining and transaction processing, are associated with significant energy consumption, leading to serious global climate issues. While the innovative financial capabilities and convenience offered by blockchain technology and cryptocurrencies are undeniable, the substantial electricity consumption and resulting environmental impacts cannot be overlooked. For example, the energy consumption of Bitcoin, the oldest and most widely used cryptocurrency, is comparable to the annual energy usage of small countries such as Denmark, Ireland, and Bangladesh \cite{kufeouglu2019bitcoin}. Additionally, Bitcoin's energy consumption contributes to significant carbon dioxide (\(CO_{2}\)) emissions; between January 1, 2016, and June 30, 2018, the Bitcoin blockchain was estimated to be responsible for up to 13 million metric tons of \(CO_{2}\) emissions. Although estimates vary, they emphasize the considerable energy demands of the network and its consequential environmental impacts, raising critical climate concerns \cite{jiang2021policy}.

To address the environmental challenges posed by blockchain technology, the Ethereum Merge represents a significant step forward, particularly concerning nonfungible tokens (NFTs). The Ethereum Merge transitioned the Ethereum blockchain from a proof-of-work (PoW) to a proof-of-stake (PoS) consensus mechanism, which drastically reduced the energy consumption required for transaction processing \cite{lal2023climate}. Furthermore, the integration of unused renewable energy sources—such as wind, solar, and hydroelectric power—into crypto mining operations presents an opportunity to replace carbon-based fuels and meet the growing demand for blockchain technology. By leveraging these underutilized energy sources and incorporating green hydrogen into crypto mining, the cryptocurrency industry can substantially decrease its carbon footprint while optimizing existing infrastructure. This approach not only addresses energy curtailment issues but also contributes positively to environmental sustainability \cite{lal2023climate,lal2024climate}.

\subsubsection{Cryptocurrency Crime}

The proliferation of cryptocurrencies has significantly transformed financial systems, introducing both opportunities and challenges. A prominent challenge is the rise in cryptocurrency-related crimes. Our topic modeling results indicate that the pursuit of sudden wealth through cryptocurrencies can motivate individuals to engage in illicit activities. The inherent anonymity and decentralized nature of cryptocurrencies facilitate a range of illegal activities, including money laundering, drug trafficking, and ransomware attacks \cite{kethineni2020rise}. The convenience theory suggests that the perceived ease and benefits of using cryptocurrencies substantially contribute to the prevalence of these crimes \cite{nolasco2021convenience}. As digital currencies continue to evolve, it is expected that new forms of financial crime will emerge, posing additional threats to global financial security \cite{trozze2022cryptocurrencies}.

Moreover, cryptocurrencies not only serve as tools for committing cybercrimes but also become prime targets for attacks, particularly due to vulnerabilities in exchanges and wallets \cite{reddy2018cryptocurrency}. Addressing these challenges requires the development of robust regulatory frameworks and the implementation of advanced technological solutions to mitigate risks and protect the integrity of financial systems. Effective collaboration among policymakers, law enforcement agencies, and technology developers is essential to create a secure and transparent environment for cryptocurrency transactions, balancing innovation with security.

\section{Conclusion}
\label{sec:conclusion}
In this study, we analyze the imbalanced patterns of public attention and sentiment toward decentralized finance across different countries, leading to four key findings.

First, our spatial aggregation map reveals distinct global imbalances in public attention and sentiment towards decentralized finance, indicating that individuals in certain countries are more active in discussing decentralized finance on social media platforms such as Twitter.

Second, the global disparities in public attention and sentiment towards decentralized finance can be statistically correlated with economic factors, including GDP per capita. Economically advanced countries are more likely to adopt and engage with new technologies, integrating them into daily discussions.

Third, our sentiment and topic analysis highlights both differences and similarities in the topics related to decentralized finance across three categories of countries. While all groups express positive sentiments towards decentralized finance, the lowest-GDP countries exhibit the strongest positive emotions. However, in these countries, cryptocurrency is often perceived as a means to achieve sudden wealth rather than as an innovative financial tool, a perspective more common in the most developed nations.
\par
For future research, we consider the potential improvements in the following aspects:
\par
\textbf{Research Question:} While extensive research has examined the relationship between blockchain and a sustainable economy~\cite{liu2021blockchain, wu2022consortium}, this study uniquely contributes to the intersection of geospatial analysis and the blockchain ecosystem. Given that geospatial locations are strongly correlated with cultural~\cite{inglehart2005modernization}, economic, and regulatory contexts, future research should explore how cultural factors influence public engagement with blockchain technologies and decentralized finance (DeFi). Incorporating cultural dimensions~\cite{graham2022corporate}—such as societal trust in technology, financial behaviors, and attitudes toward decentralization—could provide deeper insights into regional variations in blockchain adoption. Additionally, future studies could extend this analysis to other pressing social and economic issues, including sustainability (ESG factors), unemployment, inflation, and poverty reduction, by integrating additional data sources such as remote sensing and World Bank datasets. A more comprehensive understanding of these interconnections would enhance the development of blockchain-based solutions tailored to diverse geographic and cultural contexts, ultimately improving their accessibility, effectiveness, and long-term sustainability.
\par
\textbf{Methodology:} Although LDA remains a foundational method for topic modeling, integrating more advanced NLP techniques such as Large Language Models (LLMs) can enhance our approach. These models have proven effective in decoding social sentiment within decentralized autonomous organizations (DAOs) \cite{quan2023decoding} and in various sentiment analysis tasks \cite{zhang2023sentiment}. In response to the shutdown of Twitter's API, alternative open-source platforms like Discord \cite{quan2023decoding}, Reddit \cite{8999092}, and web news scrapers \cite{fu2024dam} offer viable data sources for studying community sentiments. However, these alternatives are not a panacea for data access issues, as they are centralized and could be shut down at any time. Future research should focus on developing decentralized social media platforms on blockchain infrastructures, which would democratize data access and ownership, thereby securing data for future research and development. Moreover, it is essential to incorporate geotagged data from Chinese social media platforms such as Weibo to obtain a more accurate depiction of China's public attention and sentiment toward decentralized finance. Integrating data from both Weibo and Twitter provides a comprehensive and continuous stream of global social media data \cite{liu2022assessing}. 

Last but not least, our interdisciplinary research transcends traditional boundaries, encompassing fields such as FinTech, geography, and programming. For example, the creation of geospatial maps has traditionally relied heavily on coding expertise, which may not be accessible to all researchers. To lower the barriers to advanced techniques, enhance inclusivity, and serve the public interest, we introduce KNIME, an innovative platform that eliminates the need for coding skills. With KNIME, researchers can operate in a non-programming environment to produce geospatial maps and conduct other geospatial analyses for global sentiments (see Fig. \ref{KNIME} in Appendix \ref{appendix:KNIME}).

\balance
\bibliographystyle{IEEEtran}
\bibliography{Bibliography}
\newpage
\appendix
\onecolumn
\section{DICTIONARY}

\label{sec: Dictionary}

\begin{figure}
\begin{landscape}
\includegraphics[width=\textwidth]{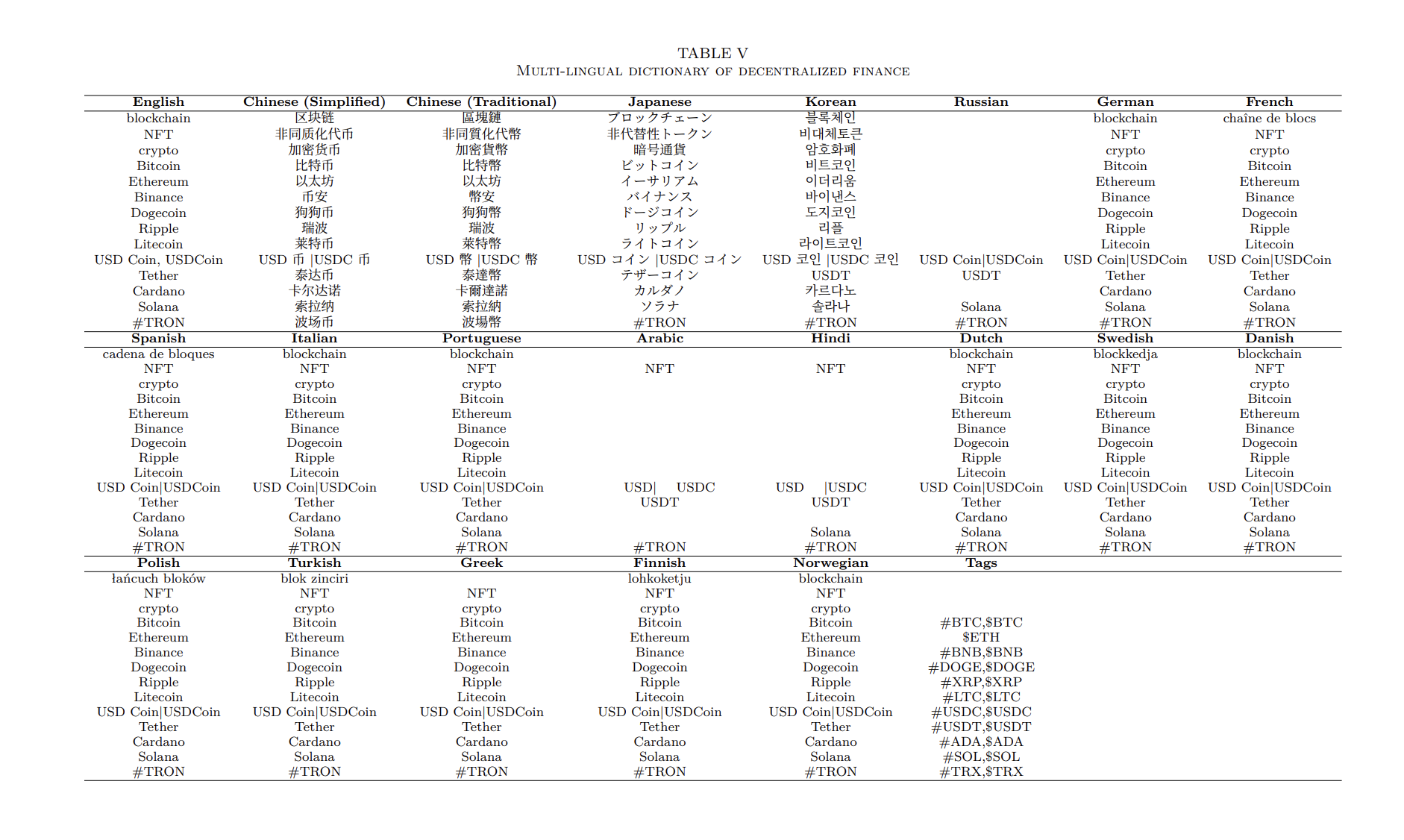}
\label{tab:dictionary}
\end{landscape}
\end{figure}

\section{DATA}
\label{sec: Data}
\begin{table}[!h]
\centering
\caption{Sample of the geo-tagged tweets}
\label{tab:sample}
\begin{tblr}{
  width = \linewidth,
  colspec = {Q[83]Q[656]Q[50]Q[79]Q[73]},
  row{1} = {c},
  rows = {m},
  cell{2}{1} = {c},
  cell{2}{3} = {c},
  cell{2}{4} = {c},
  cell{2}{5} = {c},
  cell{3}{1} = {c},
  cell{3}{3} = {c},
  cell{3}{4} = {c},
  cell{3}{5} = {c},
  cell{4}{1} = {c},
  cell{4}{3} = {c},
  cell{4}{4} = {c},
  cell{4}{5} = {c},
  hline{1-2,5} = {-}{},
}
Date & Text & Language & Place & Sentiment Score\\
2021-05-22 20:03:04 & Wait, was the Spain intro subtly advertising a dog meme crypto coin? \#Eurovision & en & Israel & 0.648121\\
2021-05-22 20:04:38 & It will ruin everyone one day in Crypto currency. & en & Gandhinagar, India & 0.072699\\
2021-05-22 20:15:06 & He means that the crash has already happened, if you wanted to sell you alts and buy BTC to protect capital, you should've done that earlier, now the pain is gone, the crash already happened. Selling alts to "protect capital" makes no sense. & en & Brasília, Brazil & 0.330482
\end{tblr}
\end{table}

\section{CLUSTERING}
\label{sec: Clusters of Countries}

\begin{figure}[!htbp]
\centering
  \includegraphics[width=0.6\textwidth]{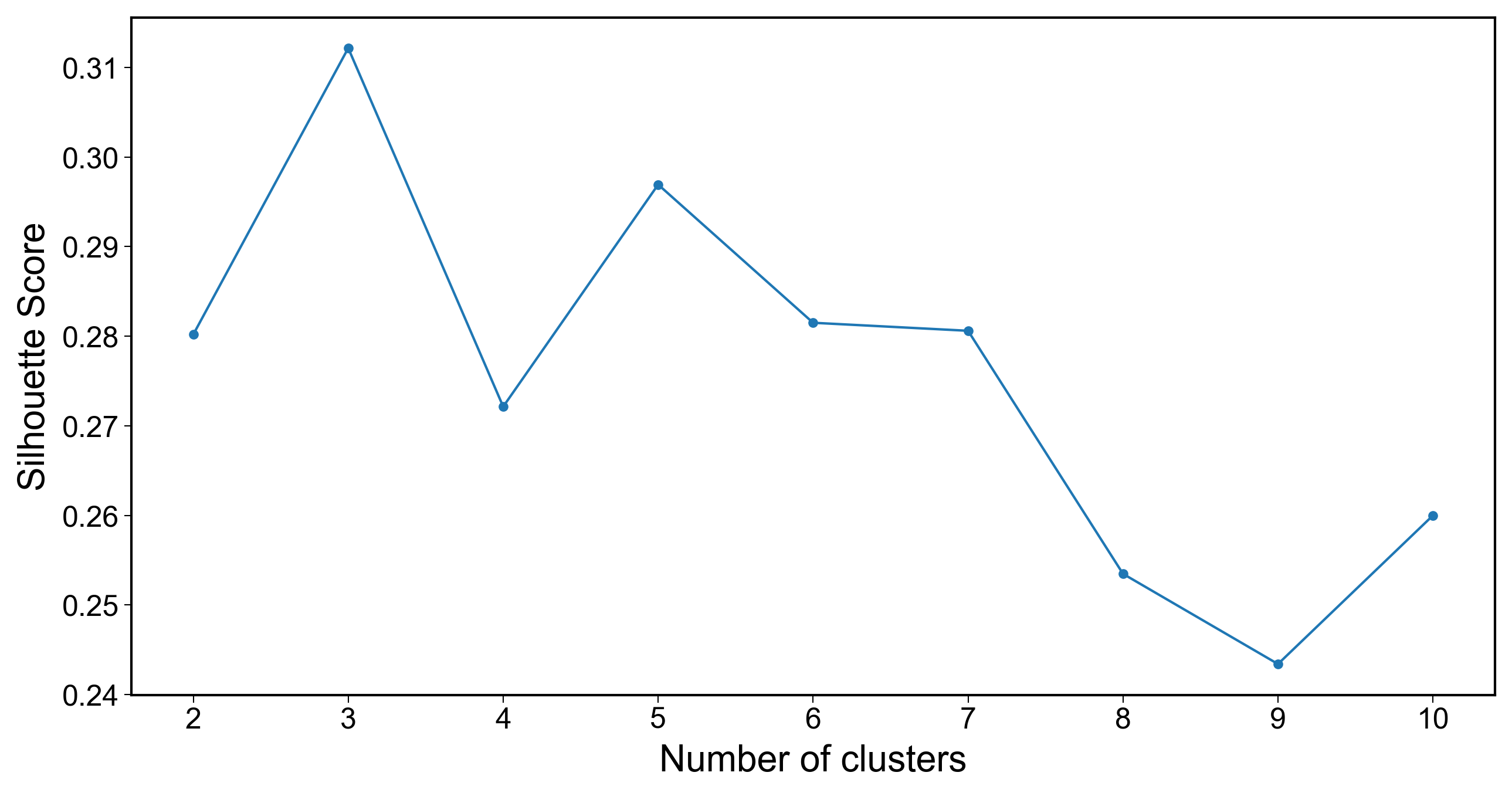}
  \caption{Silhouette score across different numbers of clusters}
  \label{fig:silhouette}
\end{figure}

\subsection{Cluster 1}
Burundi, Benin, Burkina Faso, Bangladesh, Bulgaria, Bhutan, Cameroon, Democratic Republic of the
Congo, Cuba, Gabon, Georgia, Indonesia, India, Cambodia, Laos, Sri Lanka, Morocco, Madagascar,
Myanmar, Nigeria, Pakistan, Philippines, Romania, Rwanda, Solomon Islands, Sierra Leone, Suri-
name, Syria, Togo, Ukraine, Uzbekistan, Vietnam, Samoa
\subsection{Cluster 2}
Afghanistan, Angola, Argentina, American Samoa, Azerbaijan, Belarus, Bolivia, Brazil, Botswana, Cen-
tral African Republic, Republic of Congo, Colombia, Djibouti, Dominican Republic, Algeria, Ecuador,
Egypt, Ethiopia, Fiji, Ghana, Guinea, Gambia, Equatorial Guinea, Guatemala, Guyana, Honduras, Haiti,
Iraq, Jordan, Kenya, Kyrgyzstan, Liberia, Mexico, Mali, Mongolia, Mozambique, Mauritania, Malawi,
Malaysia, Namibia, Niger, Nicaragua, Nepal, Oman, Peru, Papua New Guinea, Paraguay, Sudan, South
Sudan, Senegal, Somalia, Swaziland, Chad, Tajikistan, Tunisia, Uganda, Uruguay, Vanuatu, Yemen,
South Africa, Zambia, Zimbabwe.
\subsection{Cluster 3}
Albania, Australia, Austria, Belgium, Bosnia and Herzegovina, Brunei, Canada, Switzerland, Chile,
China, Costa Rica, Czech Republic, Germany, Denmark, Spain, Estonia, Finland, France, Faroe Islands,
United Kingdom, Greece, Croatia, Hungary, Isle of Man, Iran, Iceland, Italy, Japan, Kazakhstan, South
Korea, Lebanon, Lithuania, Luxembourg, Latvia, New Caledonia, Netherlands, Norway, New Zealand,
Panama, Poland, Portugal, Russia, El Salvador, Slovakia, Slovenia, Sweden, Thailand, Turkey, United
States of America, Venezuela

\clearpage
\section{TOPIC MODELLING}
\label{appendix:topic_modelling}

\begin{table}[!ht]
\centering
\caption{Top words of different topics in diffrent clusters.}
\begin{tblr}{
  width = \linewidth,
  colspec = {Q[35]Q[270]Q[35]Q[270]Q[35]Q[270]},
  row{1} = {c},
  row{2} = {c},
  cell{1}{1} = {c=2}{0.303\linewidth},
  cell{1}{3} = {c=2}{0.307\linewidth},
  cell{1}{5} = {c=2}{0.322\linewidth},
  cell{3}{1} = {c},
  cell{3}{3} = {c},
  cell{3}{5} = {c},
  cell{4}{1} = {c},
  cell{4}{3} = {c},
  cell{4}{5} = {c},
  cell{5}{1} = {c},
  cell{5}{3} = {c},
  cell{5}{5} = {c},
  cell{6}{1} = {c},
  cell{6}{3} = {c},
  cell{6}{5} = {c},
  cell{7}{1} = {c},
  cell{7}{3} = {c},
  cell{7}{5} = {c},
  cell{8}{1} = {c},
  cell{8}{3} = {c},
  cell{8}{5} = {c},
  cell{9}{1} = {c},
  cell{9}{3} = {c},
  cell{9}{5} = {c},
  cell{10}{1} = {c},
  cell{10}{3} = {c},
  cell{10}{5} = {c},
  cell{11}{1} = {c},
  cell{11}{3} = {c},
  cell{11}{5} = {c},
  cell{12}{1} = {c},
  cell{12}{3} = {c},
  cell{12}{5} = {c},
  cell{13}{1} = {c},
  cell{13}{3} = {c},
  cell{13}{5} = {c},
  cell{14}{1} = {c},
  cell{14}{3} = {c},
  cell{14}{5} = {c},
  cell{15}{1} = {c},
  cell{15}{3} = {c},
  cell{15}{5} = {c},
  vline{1-2,4,6} = {1}{},
  vline{1,3,5,7} = {1-15}{},
}
Cluster 1 &  & Cluster 2 &  & Cluster 3 & \\
Topic & Top words & Topic & Top words & Topic & Top words\\
1 & crypto, love, time, life, know, people, tg, eth, make, u & 1 & good, walked, tmie, jesus, lord, god, walk, news, come, make & 1 & nft, nfts, just, im, mint, project, community, yes, space, want\\
2 & promote, india, crypto, na, binance, gon, world, day, make, market & 2 & buy, just, bitcoin, nft, crypto, new, art, eth, use, work & 2 & coin, token, team, k, crypto, price, bitcoin, list, create, high\\
3 & presale, join, creepy, nft, free, token, earn, crypto, link, nfts & 3 & walk, lord, mightiest, prophets, god, come, day, u, raise, great & 3 & awesome, play, join, location, vault, check, user, cool ,game, nft\\
4 & crypto, toon, bitcoin, buy, dont, like, help, k, send, make & 4 & crypto, blockchain, im, let, airdrop, way, like, time, know, people & 4 & crypto, dont, know, just, people, like, bitcoin, im, think, make\\
5 & nft, collection, just, drop, check, painting, polygon, rarity, gas, bio & 5 & dont, check, sale, cst, day, nft, crypto, im, open, use & 5 & love, link, day, just, work, late, happy, job, u\\
6 & project, good, great, luck, best, thank, thanks, win, team, hope & 6 & link, earn, join, sign, social, network, hey, post, today, free & 6 & good, morning, u, project, say, great, strength, day, love, im\\
 &  & 7 & come, crypto, nft, collection, u, love, year, world, day, real & 7 & gm, ripple, nice, inside, xrp, day, like, effect, crypto, today\\
 &  & 8 & nft, thank, project, like, ethiopia, bitcoin, thanks, just, amazing, new & 8 & lets, na, nft, welcome, gon, x, just, congrats, make, im\\
 &  & 9 & binance, crypto, say, like, know, want, work, people, make, bitcoin & 9 & time, great, crypto, make, blockchain, year, like, week, world, good\\
 &  & 10 & today, join, space, public, pm, twitter, sleep, open, dont, follow & 10 & win, follow, today, dogecoin, like, just, post, free, hey, join\\
 &  &  &  & 11 & nft, new, collection, eth, art, drop, piece, mushroom, available, artist\\
 &  &  &  & 12 & thanks, thank, im, guy, buy, help, like, hi, country, follow\\
 &  &  &  & 13 & {bitcoin, crypto, buy, money, make, \\need, use, send, day, invest}
\end{tblr}
\end{table}

\begin{figure}[!h]
\centering
  \includegraphics[width=0.6\textwidth]{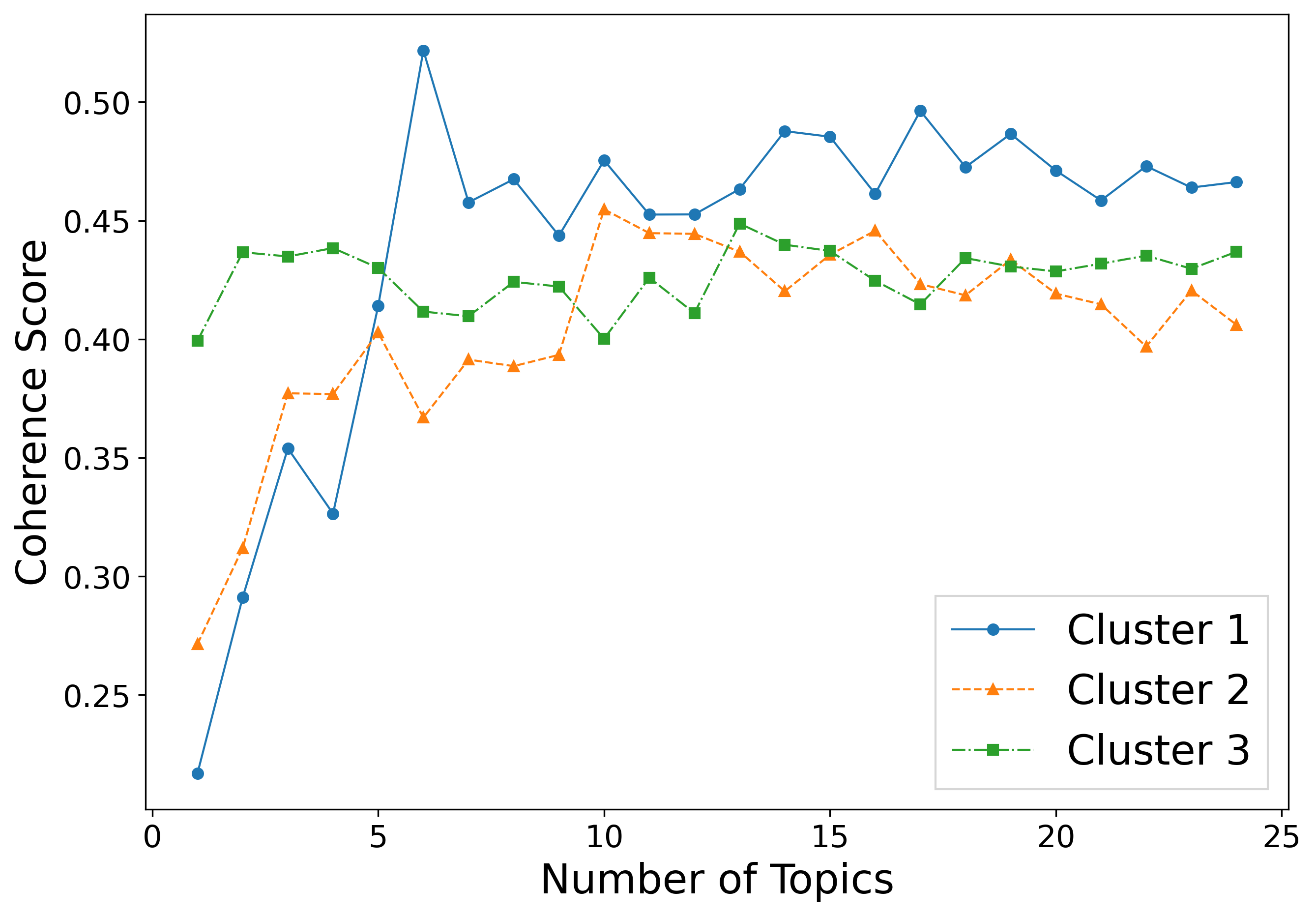}
  \caption{Topic model coherence}
  \label{fig:coherence}
\end{figure}

\section{Robuestness Checks}
\label{appendix:robuestness_checks}
\begin{figure}[!ht]
\centering
  \includegraphics[width=0.9\textwidth]{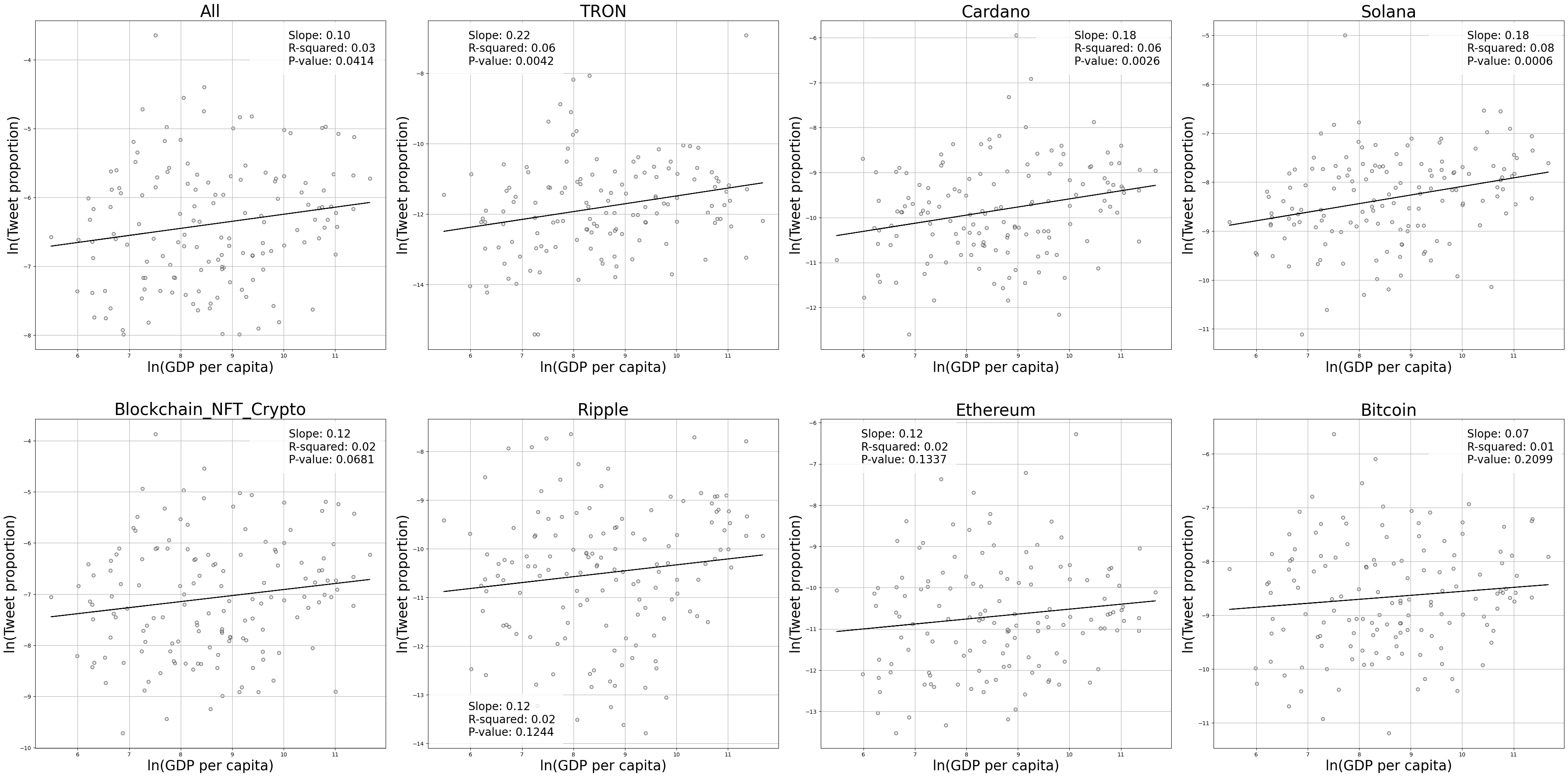}
  \caption{Linear regressions between GDP per capita and proportion of tweets related to specific topics showing significant positive correlations}
\end{figure}

\begin{figure}[!h]
\centering
  \includegraphics[width=0.75\textwidth]{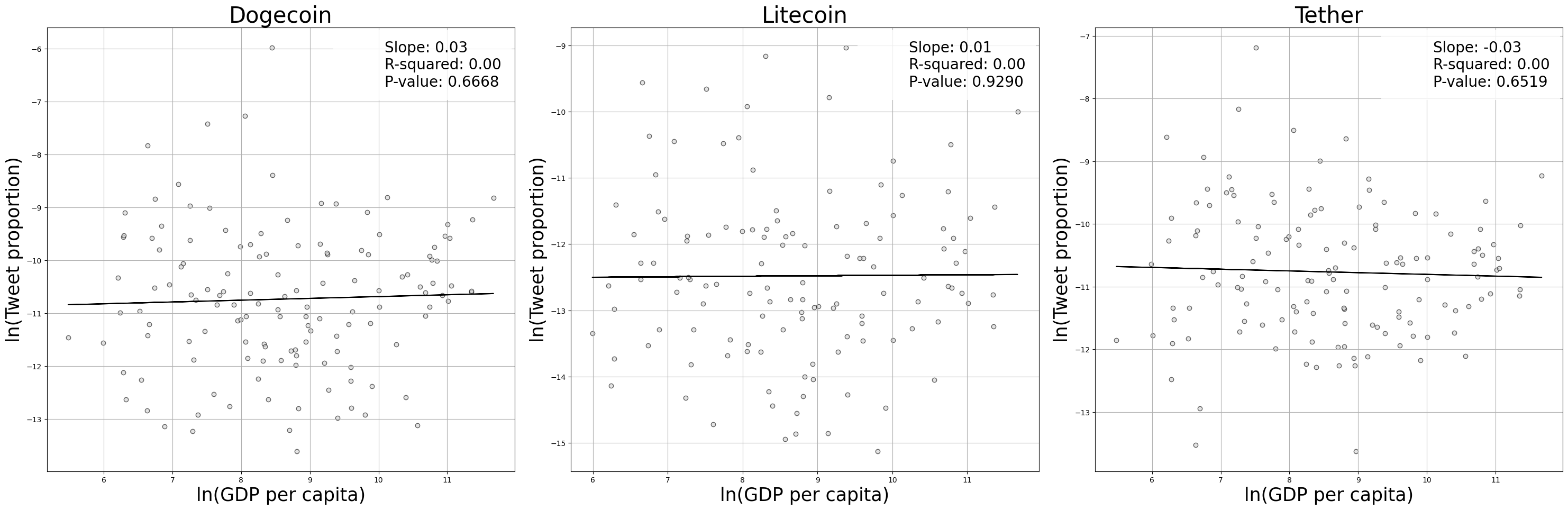}
  \caption{Linear regressions between GDP per capita and proportion of tweets related to specific topics showing insignificant correlations}
\end{figure}

\begin{figure}[!h]
\centering
  \includegraphics[width=0.5\textwidth]{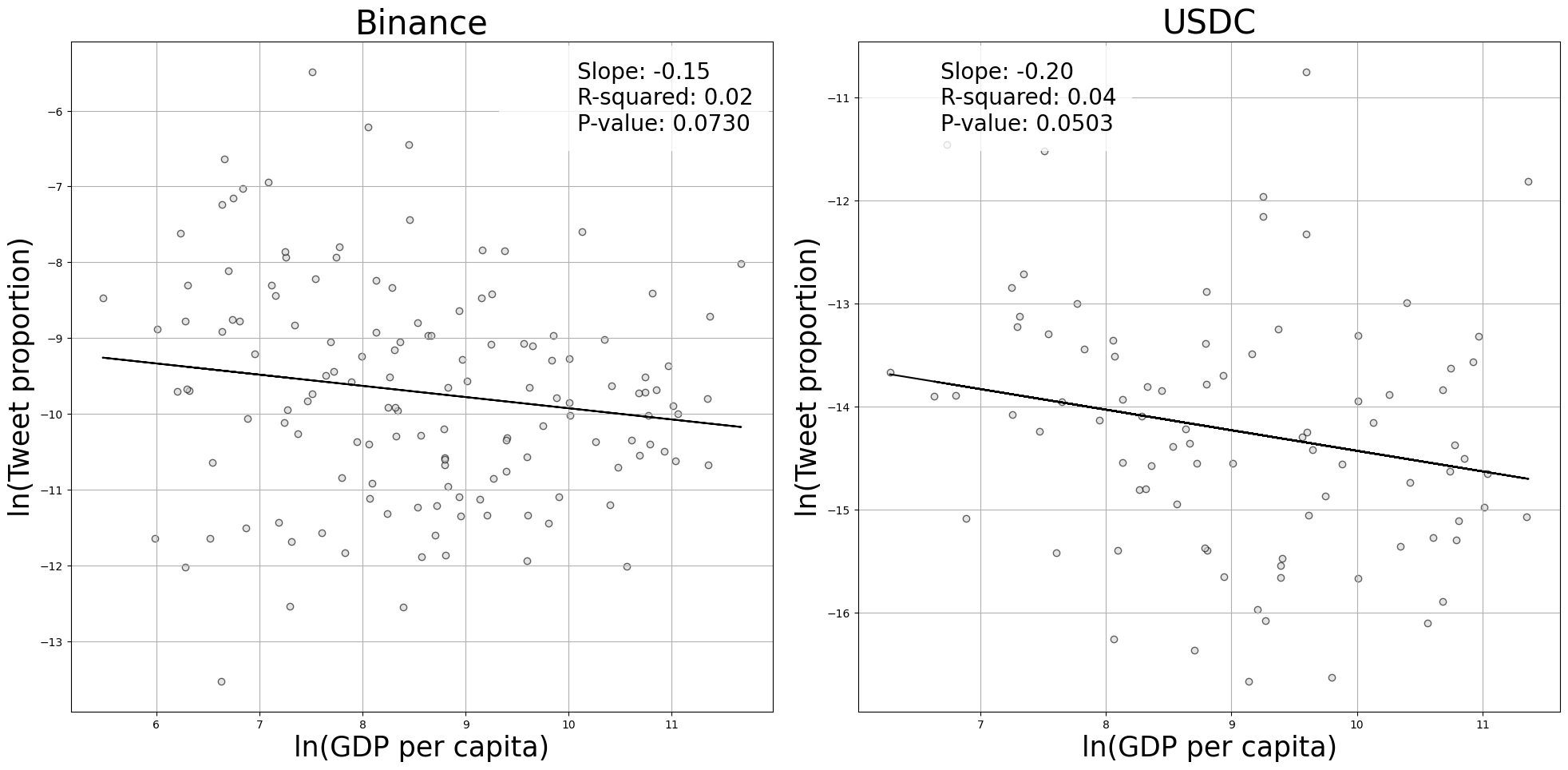}
  \caption{Linear regressions between GDP per capita and proportion of tweets related to specific topics showing significant negative correlations}
\end{figure}

\begin{figure}[!ht]
\centering
  \includegraphics[width=\textwidth]{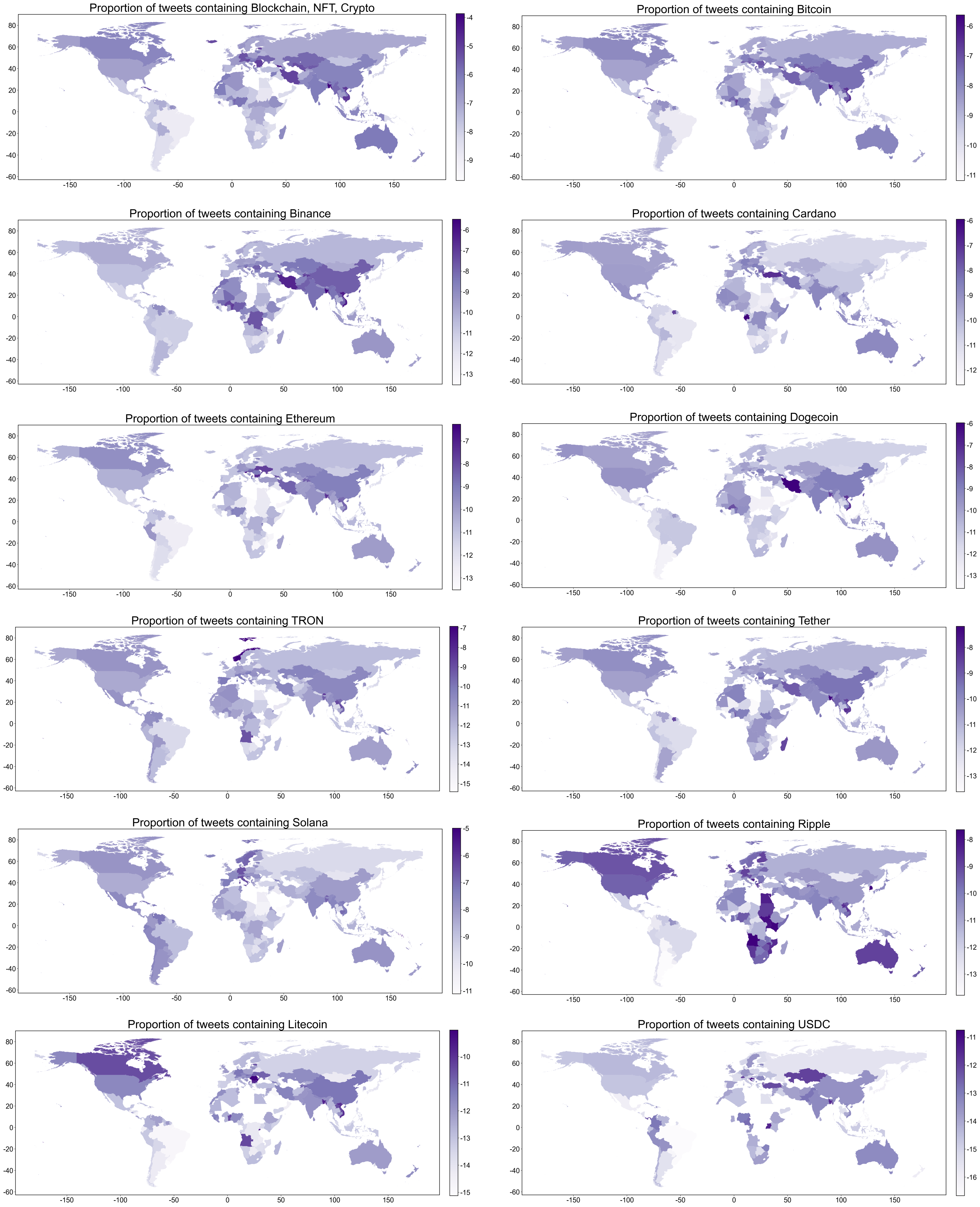}
  \caption{Proportions of tweets related to different topics across countries}
\end{figure}

\begin{figure}[!ht]
\centering
  \includegraphics[width=\textwidth]{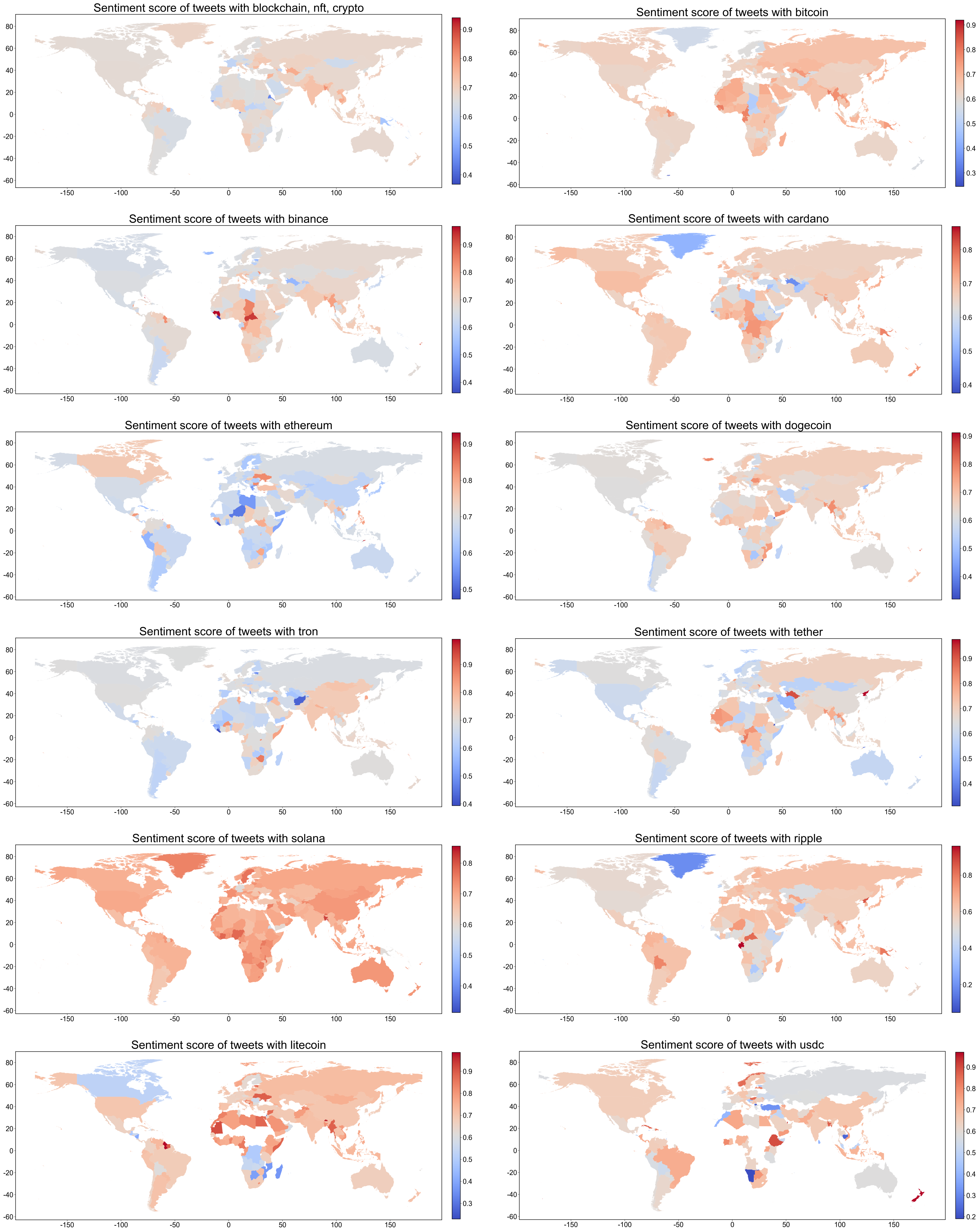}
  \caption{Sentiment scores of tweets related to different topics across countries}
\end{figure}

\clearpage

\section{Demographics}
\label{appendix:demographics}

\begin{figure}[!ht]
\centering
  \includegraphics[width=0.75\textwidth]{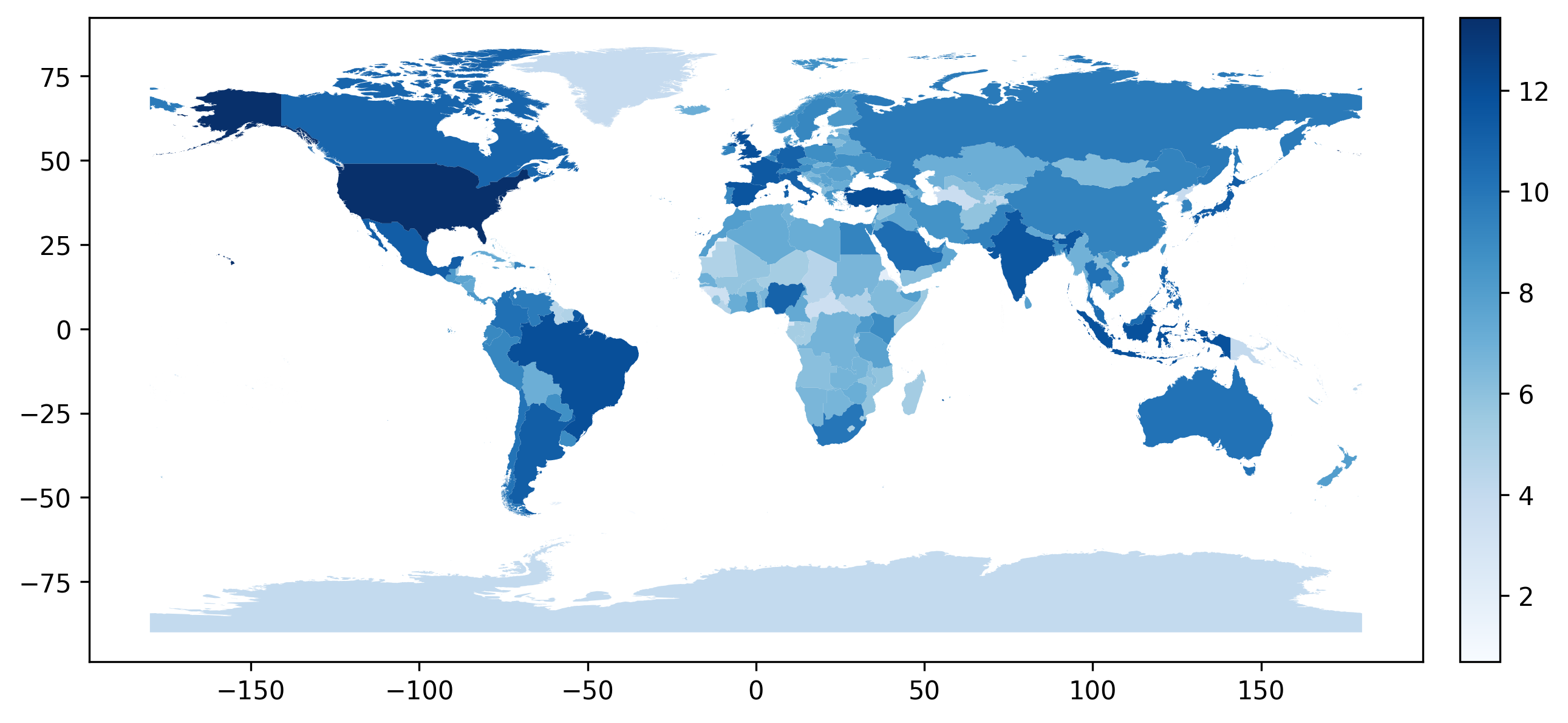}
  \caption{Logged number of users posting DeFi-related tweets across countries}
  \label{map-user}
\end{figure}

\begin{figure}[!ht]
\centering
  \includegraphics[width=0.7\textwidth]{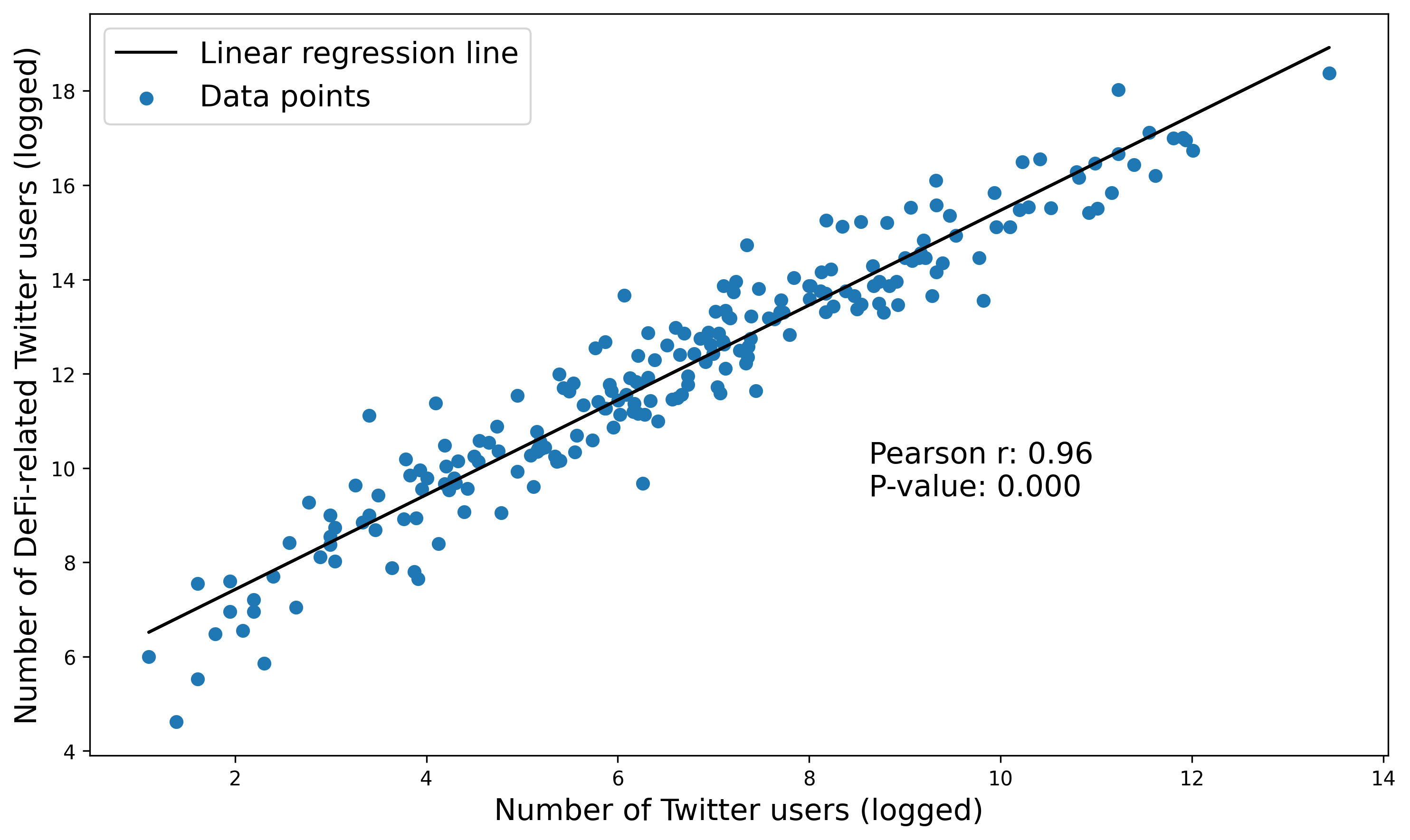}
  \caption{Linear regressions between total Twitter users and DeFi-related Twitter users}
  \label{linear-user}
\end{figure}

We calculated the number of Twitter users posting DeFi-related tweets across countries (Figure \ref{map-user}). Then we conducted Pearson correlation analysis between the number of total Twitter users \cite{WorldPopulationReview2023} and Twitter users posting DeFi-related tweets. The result (Figure \ref{linear-user}) indicates that there is a very significant positive correlation (Pearson correlation = 0.96, P-value < 0.001) between them, showing that the bias in our DeFi-related tweets dataset is essentially consistent with the inherent bias of Twitter itself.

\clearpage

\section{KNIME Workflow}
\label{appendix:KNIME}
\begin{figure}[!ht]
\centering
  \includegraphics[width=1\textwidth]{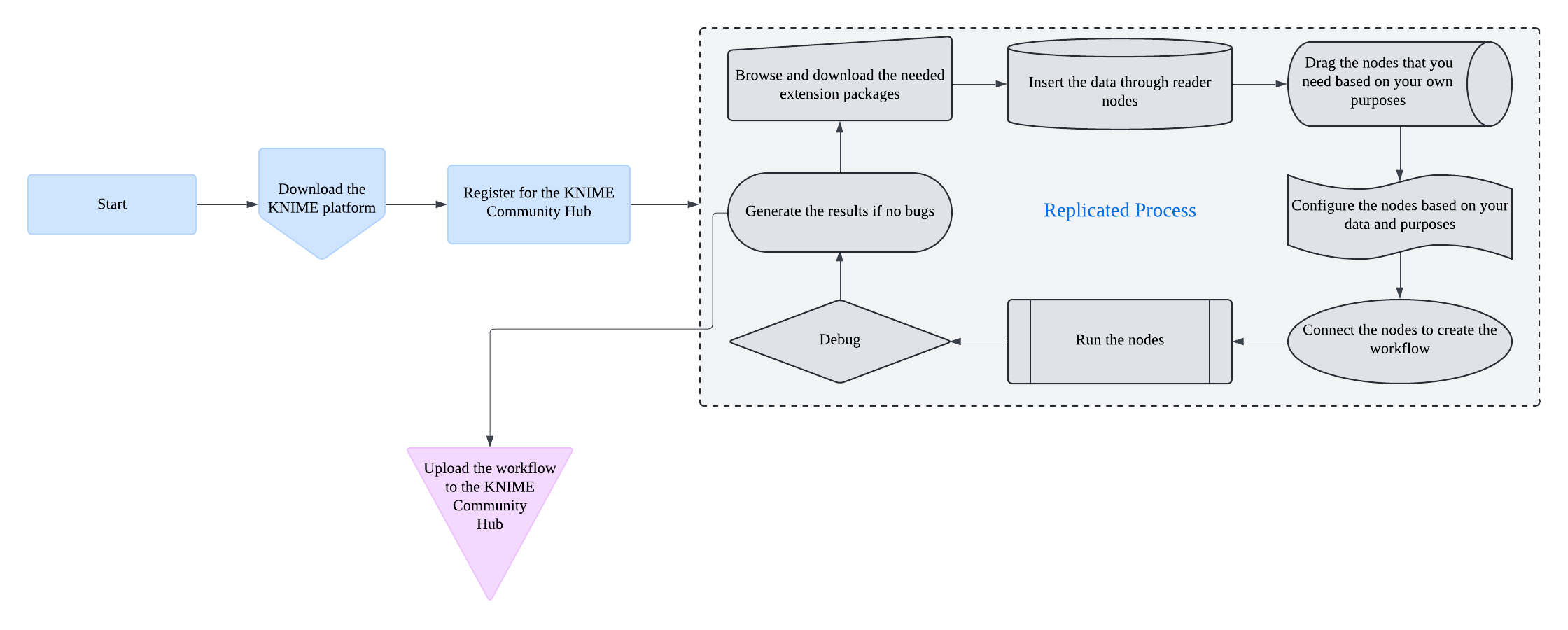}
  \caption{Process of Creating KNIME Workflow}
  \label{KNIME}
\end{figure}

\begin{figure}[!ht]
\centering
  \includegraphics[width=1\textwidth]{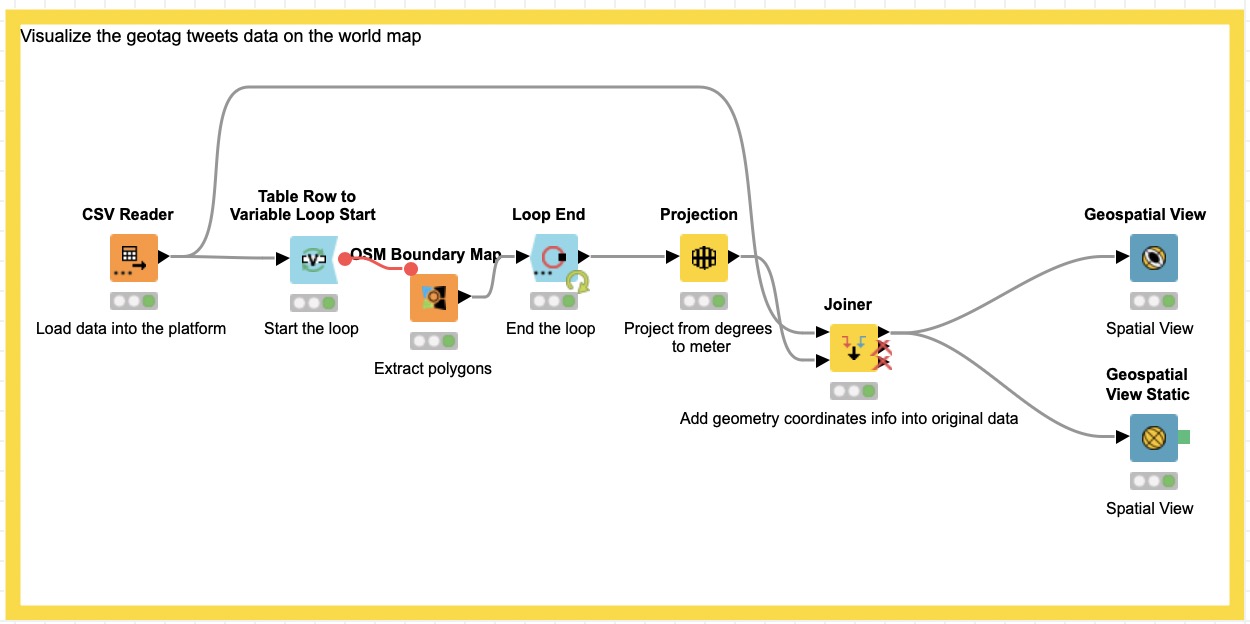}
  \caption{Complete KNIME Workflow of the Geospatial Map}
  \label{KNIME_workflow}
\end{figure}

In addition to using programming software to graph the geospatial map, we can offer a no-code solution through the KNIME platform. The generalized process of creating a workflow and drawing a map in KNIME (Figure \ref{KNIME}) is relatively straightforward and requires no coding skills. The complete KNIME workflow for generating a spatial map is depicted in Figure \ref{KNIME_workflow}. A detailed tutorial on creating and configuring this workflow is available on GitHub: \url{https://github.com/Yifanli1103/Geospatial-Map---KNIME}.

\end{document}